\documentclass[hidelinks]{article}
\usepackage[utf8]{inputenc}

\usepackage{todonotes}
\setuptodonotes{color=blue!33}
\usepackage[textwidth=450pt]{geometry} 
\usepackage{physics}
\usepackage[shortlabels]{enumitem}
\usepackage{graphicx}
\usepackage{comment}
\usepackage{multirow}
\usepackage{float}
\usepackage[
  colorlinks,
  linkcolor = blue,
  citecolor = blue,
  urlcolor = blue]{hyperref}
\usepackage{amsthm}
\usepackage[noabbrev,nameinlink]{cleveref} 
\usepackage{caption, subcaption}
\usepackage{amssymb}
\usepackage{appendix}
\usepackage{mdframed}
\usepackage{authblk}
\usepackage{mathrsfs}
\usepackage[normalem]{ulem} 
\usepackage{amsmath} 
\usepackage{wrapfig}

\usepackage{mleftright} 
\mleftright

\usepackage[
backend=biber,
style=alphabetic,
sorting=ynt
]{biblatex}
\appto{\bibsetup}{\raggedright}

\theoremstyle{plain}
\newtheorem{theorem}{Theorem}[section]
\newtheorem{prop}[theorem]{Proposition}

\newtheorem{lemma}[theorem]{Lemma}

\theoremstyle{definition}
\newtheorem{definition}[theorem]{Definition}

\crefalias{prop}{proposition}

\newcommand{\eps}{\varepsilon}
\newcommand{\pth}{p_{\text{thr}}}

\title{An efficient combination of quantum error correction and authentication}

\author[1]{Yfke Dulek}
\author[2]{Garazi Muguruza}
\author[2]{Florian Speelman}

\affil[1]{QuSoft, CWI Amsterdam, Science Park 123, 1098 XG Amsterdam, The Netherlands}
\affil[2]{QuSoft, University of Amsterdam, Science Park 904, 1098 XH Amsterdam, The
Netherlands}

\date{\normalsize\today}

\begin{document}
\maketitle

\begin{abstract}
When sending quantum information over a channel, we want to ensure that the message remains intact. Quantum error correction and quantum authentication both aim to protect (quantum) information, but approach this task from two very different directions: error-correcting codes protect against probabilistic channel noise and are meant to be very robust against small errors, while authentication codes prevent adversarial attacks and are designed to be very sensitive against any error, including small ones.

In practice, when sending an authenticated state over a noisy channel, one would have to wrap it in an error-correcting code to counterbalance the sensitivity of the underlying authentication scheme. We study the question of whether this can be done more efficiently by combining the two functionalities in a single code. To illustrate the potential of such a combination, we design the threshold code, a modification of the trap authentication code \cite{broadbent_quantum_2013} which preserves that code's authentication properties, but which is naturally robust against depolarizing channel noise. We show that the threshold code needs polylogarithmically fewer qubits to achieve the same level of security and robustness, compared to the naive composition of the trap code with any concatenated CSS code. We believe our analysis opens the door to combining more general error-correction and authentication codes, which could improve the practicality of the resulting scheme.


\end{abstract}

\clearpage
\tableofcontents
\clearpage

\section{Introduction}

Authentication is one of the most fundamental tasks of modern cryptography -- for many applications it is imperative that the integrity of data is preserved, not just against noise and random errors, but even against adversarial attacks.
Constructions for message authentication codes (MACs) underlay many important cryptographic protocols that are in constant use for secure internet communication.
We study the notion of \emph{quantum authentication}, where instead of wanting to ascertain the integrity of classical data, the data involved consists of qubits.

Starting with the seminal work of Barnum,  Crepeau, Gottesman, Smith, and Tapp~\cite{barnum_authentication_2002}, several quantum authentication codes have been proposed. In our current work, we will mostly be working with two prominent examples, namely the \emph{Clifford code} and the \emph{trap code}, not going into depth for other examples such as the polynomial code~\cite{ben-or_secure_2006} or the Auth-QFT-Auth scheme~\cite{garg_new_2017}.
The Clifford code~\cite{aharonov_interactive_2008} constructs a very effective authentication scheme, which involves attaching a number of flag qubits to the plaintext, and then scrambling the state using a random Clifford -- this turns out to be a very efficient way of guaranteeing security, and it can also be used as a building block for interactive proofs~\cite{aharonov_interactive_2008} and multi-party computation~\cite{dupuis_secure_2010,dupuis_actively_2012,dulek_secure_2020}.
The trap code~\cite{broadbent_quantum_2013,broadbent_efficient_2016} constructs a scheme, for which encoding consists of interspersing the plaintext (in an error-correcting code) with so-called traps which try to detect an adversary's attempted modifications. This code has been used to build quantum one-time programs~\cite{broadbent_quantum_2013}, zero-knowledge proofs~\cite{broadbent_zero-knowledge_2016}, and verifiable homomorphic encryption~\cite{alagic_quantum_2017}.

Multiple works have followed the first notions of security for the primitive of quantum authentication of Barnum et al.~\cite{barnum_authentication_2002}, which did not consider adversaries entangled with the encrypted message. An important requirement for authentication protocols is a composable security notion, which ensures that the scheme is secure when using it in any arbitrary environment. By using a simulator-based approach to security, several additional desirable properties to the basic functionality have been proven, such as key recycling~\cite{hayden_universal_2016,portmann_quantum_2017,garg_new_2017} or quantum ciphertext authentication~\cite{alagic_unforgeable_2018,dulek_quantum_2018}. Additionally, it is possible to study the notion of authentication in the setting of computational security~\cite{banfi_composable_2019}, including public-key versions of the primitive~\cite{alagic_can_2021}. In this work we extensively use the Abstract Cryptography framework introduced by Maurer and Renner~\cite{maurer_abstract_2011}, which views cryptography as a resource theory and has been previously applied successfully to prove security of purity testing based authentication schemes by Portmann~\cite{portmann_quantum_2017}.

Authentication is usually applied to messages that will be transmitted at some point, and such a transmission involves incurring some \emph{error} by the quantum channel which is used. The MACs present in the literature will inevitably reject whenever any error is present in the channel. However, it is possible to first encode this message in a quantum authentication code, and then wrap the result in an \emph{error-correction code} (see also e.g.~the discussion by \cite{hayden_universal_2016} and mainly~\cite{portmann_quantum_2017}). 

Observe that the primitives of quantum authentication and error correction have a conceptual overlap, in the sense that both aim to protect data against modifications. However, in practice there is a large difference in how they are built: authentication codes need to protect against any adversarial attack, and therefore often are extremely sensitive against even minor attempted modifications.
For example, if a Pauli operation would be applied to a single qubit that is part of a Clifford-code authenticated state, the encoded plaintext would be almost completely scrambled by having a random $n$-qubit Pauli operator applied to the entire plaintext. On the other hand, an error-correcting code should be robust against typical (usually low-weight) modifications of the encoded data. Given that the goals of these codes are similar, one might wonder whether this is doing `double work' in some sense, making the resulting encoded state larger than necessary.

In this work, we give evidence that this is indeed the case: We construct a code which functions both as a quantum error-correcting code and as a quantum authentication code, and which is more efficient than the naive concatenation of these functionalities would imply. In particular:
\begin{itemize}
	\item As an example of a combined code, we present the \emph{threshold code}.  Even though this is not the main goal of the current work, note that this code preserves several of the useful computational properties of the original trap code, if a CSS code is used as the underlying error-correcting code, having essentially the same encoding procedure as the trap code and only differing in the decoding.
        \item We adapt the definition of cryptographic security in the AC framework, by splitting up the correctness and security of authentication, and show that it is still composable.
	\item We show that our scheme is correct and secure, by proving that the resulting code is a good purity testing code. Because of the generality of the AC framework, the same security proof will also imply security under most other security definitions (if these do not require extra properties such as key recycling).
	\item We compare the resulting scheme to the concatenation of the two primitives separately. If we define efficiency in terms of amount of qubits needed to obtain certain correctness and security for a constant-error quantum depolarizing channel, we show how the resulting scheme is more efficient than applying the codes separately.
\end{itemize}
\section{Preliminaries}
\subsection{Notation}

The single-qubit Pauli matrices given by
\[ I=\begin{pmatrix}1&0\\0&1\end{pmatrix},\: X=\begin{pmatrix}0&1\\1&0\end{pmatrix},\: Z=\begin{pmatrix}1&0\\0&-1\end{pmatrix},\:\text{and}\: Y=\begin{pmatrix}0&-i\\i&0\end{pmatrix}=iXZ, \]
form a basis for single-qubit Pauli operations. Note that they are unitaries and any two Pauli operations either commute or anti-commute. An $n$-qubit Pauli matrix is given by $n$-fold tensor products of single-qubit Paulis, and we denote the \emph{Pauli group} of $n$-qubit Pauli matrices by
\[ \mathcal{G}_n:=\{i^k P_1\otimes P_2\otimes\cdots\otimes P_n\::\: \text{where}\: P_j\in\{I,X,Z,Y\},\: k\in [4]\}. \]
The \emph{weight} of an $n$-qubit Pauli operation, denoted $\omega(P)$, is the number of non-identity Paulis in the $n$-fold tensor product. Moreover, we denote by $\omega_X(P)$ and $\omega_Z(P)$ the number of $X$ and $Z$-Paulis respectively.

The \emph{Clifford group}, $\mathcal{C}_n$, is the group of $n$-qubit unitaries that leave the Pauli group invariant. That is, given $P\in\mathcal{G}_n$, for all $C\in\mathcal{C}_n$ we have $i^{k} CPC^\dag\in\mathcal{G}_n$, where $k\in[4]$.

The logarithm $\log$ is considered in base $2$, unless specified otherwise.

\subsection{Abstract cryptography}

In this article we follow the Abstract Cryptography (AC) framework for cryptography~\cite{maurer_abstract_2011}, which was first used to model the composable security of quantum authentication by Portmann~\cite{portmann_quantum_2017}. AC views cryptography as a resource theory: a protocols constructs an \emph{ideal resource} from a \emph{real system} by means of a \emph{simulator}. We will describe the basic concepts here, but since the relevant results will be referenced from previous works we recommend reading the works~\cite{portmann_quantum_2017, portmann_security_2021, maurer_abstract_2011} for a deeper understanding.

In an $n$-player setting, a \emph{resource} is an object with $n$ interface; allows the players to input and receive messages. We will denote resources by squares, and inputs/outputs from the interfaces by lines intersecting with the squares. If two resources $\mathcal{C}$ and $\mathcal{K}$ are available to the players, we write $\mathcal{C}||\mathcal{K}$ for the parallel composition of the resources: the resources are simultaneously accessible to the players in any arbitrary order, thus in particular, the order of composition is irrelevant and $\mathcal{C}||\mathcal{K}=\mathcal{K}||\mathcal{C}$.

A \emph{converter} models the local operations that the players can perform in their interfaces. We will denote converters by squares with rounded corners. If a converter $\sigma$ is connected to the interface $i$ of the resource $\mathcal{C}$, we write $\sigma_i\mathcal{C}$ (or equivalently $\mathcal{C}\sigma_i$). A \emph{protocol} is defined by a set of converters: one for each honest player. On the one hand, an adversary is allowed to perform any operation allowed by quantum mechanics, thus it is essential to prove security against adversaries. On the other hand, for security guarantees it is not enough to show good performance in presence of adversaries, we also need to emulate the presence of no adversary. We do this with a special type of converters, called \emph{filters}, which emulate the honest behaviour of the adversary. We call \emph{filtered resources} a pair of resource $\mathcal{C}$ and filter $\Diamond_E$, denoted $\mathcal{C}_{\Diamond}=(\mathcal{C},\Diamond_E)$. 

We can now define a composable security notion in the AC framework. Composability is an essential requirement for security as it argues that the protocol will be secure in any arbitrary environment; in particular, both against substitution and impersonation attacks, relevant for authentication.
\begin{definition}[Cryptographic security]\label{def:security}
We say that the protocol $\pi_{AB}=(\pi_A,\pi_B)$ constructs the filtered resource $\mathcal{S}_\Diamond$ from $\mathcal{C}_\#$ within $(\eps,\delta)$, denoted $\mathcal{C}_\#\xrightarrow{\pi_{AB},(\eps,\delta)}\mathcal{S}_\Diamond$, if the following two conditions hold:
    \begin{enumerate}[i)]
        \item\label{item:eps} In presence of no malicious player, the filtered resources are $\eps$-close to each other
        \[ d(\pi_{AB}\mathcal{C}_\#,{S}_\Diamond)\leq\eps. \]
        \item\label{item:delta} In the presence of an adversary, there exists a simulator $\sigma_E$, $\delta$-close to the real protocol
        \[ d(\pi_{AB}\mathcal{C},\sigma_E\mathcal{S} )\leq\delta. \]
    \end{enumerate}
    Here the distance $d$ is the supremum over the set of all possible distinguishers allowed by quantum mechanics. If the filtered resources $\mathcal{S}m_\Diamond$ and $\mathcal{C}_\#$ are clear from the context, we say that $\pi_{AB}$ is $(\eps,\delta)$-secure, or $\eps$-correct and $\delta$-secure.
\end{definition}

We differ from the original definition of cryptographic security in~\cite{maurer_abstract_2011}, where security is defined as the maximum of the two values $\eps$ and $\delta$, because these parameters have independent meanings that are interesting to study separately. The $\eps$ in \cref{item:eps} refers to the \emph{correctness} of the protocol. That is, the probability that the protocol running on a noisy channel without adversary will be distinguishable from an ideal channel. The $\delta$ in \cref{item:delta} is the usual \emph{security} in presence of an adversary. Although we might want to consider equal correctness and security in certain scenarios, splitting these two parameters allows us to revisit the proofs from Portmann and understand composability of authentication and error-correction in terms of cryptographic security parameters. For example, authentication protocols considered in the literature are not correct in presence of noisy, i.e.\ they will always reject with high probability, unless they are wrapped in error-correcting codes, which are correct but not necessarily secure.

The split parameters provide a more refined composable security notion.
\begin{theorem}[Serial composition security]\label{thm:composition}
Let the protocols $\pi$ and $\pi'$ construct $\mathcal{S}_\Diamond$ from $\mathcal{R}_\#$ and $\mathcal{T}_\square$ from $\mathcal{S}_\Diamond$ within $(\eps,\delta)$ and $(\eps',\delta')$ respectively, i.e.
\[ \mathcal{R}_\#\xrightarrow{\pi,(\eps,\delta)}\mathcal{S}_\Diamond\quad\text{and}\quad\mathcal{S}_\Diamond\xrightarrow{\pi',(\eps',\delta')}\mathcal{T}_\square. \]
Then the serial composition $\pi'\pi$ constructs $\mathcal{T}_\square$ from $\mathcal{R}_\#$ within $(\eps+\eps',\delta+ \delta')$,
\[ \mathcal{R}_\#\xrightarrow{\pi'\pi,(\eps+\eps',\delta+\delta')}\mathcal{T}_\square. \]
    \begin{proof}
    The statement follows directly from the triangle inequality. For $\eps$-correctness we have that
    \[d(\pi'\pi\mathcal{R}_\#,\mathcal{T}_\square) \leq d(\pi'\pi\mathcal{R}_\#,\pi'\mathcal{S}_\Diamond)+d(\pi'\mathcal{S}_\Diamond,\mathcal{T}_\square)
    \leq d(\pi\mathcal{R}_\#,\mathcal{S}_\Diamond)+\eps'\leq\eps+\eps'. \]
    Similarly for $\delta$-security, the composed converter $\sigma'\sigma$ is a converter for the composition since
    \[ d(\pi'\pi\mathcal{R},\sigma'\sigma\mathcal{T}) \leq d(\pi'\pi\mathcal{R},\pi'\sigma\mathcal{S} )+d(\pi'\sigma\mathcal{S},\sigma'\sigma\mathcal{T})
    \leq d(\pi\mathcal{R},\sigma\mathcal{S})+d(\pi'\mathcal{S},\sigma'\mathcal{T}) \leq\delta+\delta', \]
    where we used commutativity of converters $\alpha\beta\mathcal{C}=\beta\alpha \mathcal{C}$ and the pseudo-metric property $d(\alpha\mathcal{C},\alpha\mathcal{C}')\leq d(\mathcal{C},\mathcal{C}')$, see \cite{modersheim_theory_2012}.
    \end{proof}
\end{theorem}

\subsection{Quantum error correction}\label{sec:error}

Since quantum information is very sensitive to errors and noise from the environment; quantum error correction is developed as a tool to protect data against errors. A $[[n,k,d]]$ quantum error-correcting code (QECC) is an encoding of $k$ `logical qubits' (which we wish to protect from errors) into a codeword consisting of $n$ `physical qubits' (auxiliary qubits), with $n>k$. The distance $d$ is the minimum weight of a Pauli $P$ to turn one valid codeword into another.

After the encoded information is subjected to noise, we perform a collective measurement on the~$n$ qubits which will enable us to diagnose the type of error that occurred, called the error syndrome. Afterwards, error decoding or recovery is performed, to return to the original state of the code. We say that a $[[n,k,d]]$ QECC can correct $t$ errors if recovery is successful for any superoperator with support on the set of Pauli operators of weight up to $t$. In any case, we assume that we can always decode, possibly to a different state than the input if more than $t$ errors are present. Moreover, sometimes we are satisfied just with knowing if an error has occurred, without the need to reverse it. We call this the error-detection property of the code. In fact, a QECC with distance $d$ can correct $t=(d-1)/2$ errors. For a more in detail analysis we refer the refer the reader to standard literature in error correction~\cite{nielsen_quantum_2010, preskill_lecture_1999}.


A general way to construct quantum authentication codes is by using purity-testing error-correcting codes~\cite{barnum_authentication_2002}. For simplicity of notation, we will focus on purity-testing codes based on stabilizer codes. Stabilizer codes are best studied with the stabilizer formalism, developed by Gottesman~\cite{gottesman_class_1996}. In short, they allow us to describe quantum states in terms of the operators stabilizing them instead of working with the state itself, by means of group theory techniques for the Pauli group.

Any two elements of the Pauli group $\mathcal{G}_n$ either commute or anti-commute and square to $\pm I$, which we will use to describe codewords. Given an abelian subgroup $S$ of $\mathcal{G}_n$, we define the \emph{stabilizer code} $V_S$ to be the stable states under the action of elements of $S$. That is, 
\[ V_S:=\{ \ket{\psi}\colon M\ket{\psi}=\ket{\psi},\quad M\in S \}. \]
Let us denote by $S_1,\ldots, S_m$ the generators of the stabilizer group $S=\langle S_1,\ldots,S_m \rangle$.
Since any Pauli error $P\in\mathcal{G}_n$ either commutes or anti-commutes with each element of the generator, we can define the vector $s_P=(s_{1,P},\ldots,s_{m,P})$ such that $s_j=0$ if $P_j$ commutes with $S_j$, and $s_j=1$ if it anti-commutes. Therefore,
\[ S_jP\ket{\psi}=(-1)^{s_{j,P}} P S_i\ket{\psi}= (-1)^{s_{j,P}} P\ket{\psi}, \quad\text{for all}\:\ket{\psi}\in V_S.\]
We  call the vector $s$ the \emph{syndrome} of the error-correcting code. Errors with non-zero syndrome for some element in the stabilizer $M\in S$ can be detected by the QECC -- i.e.\ the ones that anti-commute with some element of the stabilizer. However, commuting errors are undetectable, and will change the code whenever they are not part of the stabilizer. If we denote by $S^\bot$ the set of Paulis that commute with the stabilizer, i.e.
\[ S^\bot:=\{P\in\mathcal{G}_n\colon PM=MP\:\:\text{for all}\: M\in S\}, \]
then the set of undetectable errors that change the data non-trivially is $S^\bot\setminus S$.  


Purity testing codes are exactly the stabilizer codes that detect any non-trivial Pauli attack with high probability. This property makes them extremely well suited for constructing authentication schemes as we will see in~\Cref{sec:authentication}.

\begin{definition}
    A set $\{V_k\}_{k\in \mathcal{K}}$ of stabilizer codes, each with respective stabilizer subgroup $S_k$, is $\eps$-\emph{purity testing} if, when the code is selected uniformly at random, the probability of any Pauli error $P\in\mathcal{G}_n$ acting non-trivially on the data and not being detected is upper bounded by $\eps$. That is,
    \[ \Pr_{k\in\mathcal{K}}\left(P\in S^\bot_k\setminus S_k\right)\leq\eps. \]
\end{definition}


Another class of interesting error-correcting codes is the \emph{concatenated} codes. Given a $[[n, 1 ,d]]$ QECC, we can recursively encode each encoded qubit in $n$ physical qubits, which can be encoded again such that each layer $L$ of concatenation is a $[[n^L, 1, d^L]]$ QECC, see \cite{preskill_lecture_1999}. Although the ratio of correctable errors tends to zero with the number of concatenations,
\[ \frac{t^{(L)}}{n^L}=\frac{d^L-1}{2n^L}\to 0, \]
the probability of failed recovery drops double exponentially, in other words, errors have to be spread in a very conspiratorial fashion to undermine recovery. 

If the failure probability of each qubit in the lowest level (physical qubits) is $p$, assuming that the qubits are subjected to independent and identically distributed errors, we can model the error by random variables $X_1,\ldots,X_{n^L}$ such that $X_j\in\{X,Y,Z\}$ with probability $p$ and $X_j=I$ with probability $1-p$. However, if we want to study the probability of the error correction failing to decode properly, we have to attend to the layered decoding as well. We can construct a pyramid of dependent variables such that for each layer $l=0,\ldots,L-1$ we have

\begin{minipage}{0.38\linewidth}
    \begin{equation*}\begin{split}
    X_1^{(l)}&:=X^{(l+1)}_1\otimes \cdots\otimes X^{(l+1)}_n,\\
    X_2^{(l)}&:=X^{(l+1)}_{n+1}\otimes \cdots\otimes X^{(l+1)}_{2n},\\
    \vdots&\\
    X_{n^{l}}^{(l)}&:=X^{(l+1)}_{n^l-n+1}\otimes \cdots\otimes X^{(l+1)}_{n^l},
    \end{split}\end{equation*}
\end{minipage}%
\hfill%
\begin{minipage}{0.58\linewidth}
    \centering\begin{tikzpicture}[x=0.75pt,y=0.75pt,yscale=-1,xscale=1]

\draw    (73,100.49) -- (107.59,118.2) ;
\draw [shift={(109.38,119.11)}, rotate = 207.11] [color={rgb, 255:red, 0; green, 0; blue, 0 }  ][line width=0.75]    (10.93,-3.29) .. controls (6.95,-1.4) and (3.31,-0.3) .. (0,0) .. controls (3.31,0.3) and (6.95,1.4) .. (10.93,3.29)   ;
\draw    (73,100.49) -- (107.63,81.09) ;
\draw [shift={(109.38,80.11)}, rotate = 150.74] [color={rgb, 255:red, 0; green, 0; blue, 0 }  ][line width=0.75]    (10.93,-3.29) .. controls (6.95,-1.4) and (3.31,-0.3) .. (0,0) .. controls (3.31,0.3) and (6.95,1.4) .. (10.93,3.29)   ;
\draw    (73,100.49) -- (107.38,100.13) ;
\draw [shift={(109.38,100.11)}, rotate = 179.41] [color={rgb, 255:red, 0; green, 0; blue, 0 }  ][line width=0.75]    (10.93,-3.29) .. controls (6.95,-1.4) and (3.31,-0.3) .. (0,0) .. controls (3.31,0.3) and (6.95,1.4) .. (10.93,3.29)   ;

\draw    (146.5,76.49) -- (181.09,94.2) ;
\draw [shift={(182.88,95.11)}, rotate = 207.11] [color={rgb, 255:red, 0; green, 0; blue, 0 }  ][line width=0.75]    (10.93,-3.29) .. controls (6.95,-1.4) and (3.31,-0.3) .. (0,0) .. controls (3.31,0.3) and (6.95,1.4) .. (10.93,3.29)   ;
\draw    (146.5,76.49) -- (181.13,57.09) ;
\draw [shift={(182.88,56.11)}, rotate = 150.74] [color={rgb, 255:red, 0; green, 0; blue, 0 }  ][line width=0.75]    (10.93,-3.29) .. controls (6.95,-1.4) and (3.31,-0.3) .. (0,0) .. controls (3.31,0.3) and (6.95,1.4) .. (10.93,3.29)   ;
\draw    (146.5,76.49) -- (180.88,76.13) ;
\draw [shift={(182.88,76.11)}, rotate = 179.41] [color={rgb, 255:red, 0; green, 0; blue, 0 }  ][line width=0.75]    (10.93,-3.29) .. controls (6.95,-1.4) and (3.31,-0.3) .. (0,0) .. controls (3.31,0.3) and (6.95,1.4) .. (10.93,3.29)   ;

\draw    (147,125.49) -- (181.59,143.2) ;
\draw [shift={(183.38,144.11)}, rotate = 207.11] [color={rgb, 255:red, 0; green, 0; blue, 0 }  ][line width=0.75]    (10.93,-3.29) .. controls (6.95,-1.4) and (3.31,-0.3) .. (0,0) .. controls (3.31,0.3) and (6.95,1.4) .. (10.93,3.29)   ;
\draw    (147,125.49) -- (181.63,106.09) ;
\draw [shift={(183.38,105.11)}, rotate = 150.74] [color={rgb, 255:red, 0; green, 0; blue, 0 }  ][line width=0.75]    (10.93,-3.29) .. controls (6.95,-1.4) and (3.31,-0.3) .. (0,0) .. controls (3.31,0.3) and (6.95,1.4) .. (10.93,3.29)   ;
\draw    (147,125.49) -- (181.38,125.13) ;
\draw [shift={(183.38,125.11)}, rotate = 179.41] [color={rgb, 255:red, 0; green, 0; blue, 0 }  ][line width=0.75]    (10.93,-3.29) .. controls (6.95,-1.4) and (3.31,-0.3) .. (0,0) .. controls (3.31,0.3) and (6.95,1.4) .. (10.93,3.29)   ;

\draw    (210,56.49) -- (244.59,74.2) ;
\draw [shift={(246.38,75.11)}, rotate = 207.11] [color={rgb, 255:red, 0; green, 0; blue, 0 }  ][line width=0.75]    (10.93,-3.29) .. controls (6.95,-1.4) and (3.31,-0.3) .. (0,0) .. controls (3.31,0.3) and (6.95,1.4) .. (10.93,3.29)   ;
\draw    (210,56.49) -- (244.63,37.09) ;
\draw [shift={(246.38,36.11)}, rotate = 150.74] [color={rgb, 255:red, 0; green, 0; blue, 0 }  ][line width=0.75]    (10.93,-3.29) .. controls (6.95,-1.4) and (3.31,-0.3) .. (0,0) .. controls (3.31,0.3) and (6.95,1.4) .. (10.93,3.29)   ;
\draw    (210,56.49) -- (244.38,56.13) ;
\draw [shift={(246.38,56.11)}, rotate = 179.41] [color={rgb, 255:red, 0; green, 0; blue, 0 }  ][line width=0.75]    (10.93,-3.29) .. controls (6.95,-1.4) and (3.31,-0.3) .. (0,0) .. controls (3.31,0.3) and (6.95,1.4) .. (10.93,3.29)   ;

\draw    (210.5,146.99) -- (245.09,164.7) ;
\draw [shift={(246.88,165.61)}, rotate = 207.11] [color={rgb, 255:red, 0; green, 0; blue, 0 }  ][line width=0.75]    (10.93,-3.29) .. controls (6.95,-1.4) and (3.31,-0.3) .. (0,0) .. controls (3.31,0.3) and (6.95,1.4) .. (10.93,3.29)   ;
\draw    (210.5,146.99) -- (245.13,127.59) ;
\draw [shift={(246.88,126.61)}, rotate = 150.74] [color={rgb, 255:red, 0; green, 0; blue, 0 }  ][line width=0.75]    (10.93,-3.29) .. controls (6.95,-1.4) and (3.31,-0.3) .. (0,0) .. controls (3.31,0.3) and (6.95,1.4) .. (10.93,3.29)   ;
\draw    (210.5,146.99) -- (244.88,146.63) ;
\draw [shift={(246.88,146.61)}, rotate = 179.41] [color={rgb, 255:red, 0; green, 0; blue, 0 }  ][line width=0.75]    (10.93,-3.29) .. controls (6.95,-1.4) and (3.31,-0.3) .. (0,0) .. controls (3.31,0.3) and (6.95,1.4) .. (10.93,3.29)   ;

\draw (6,90.9) node [anchor=north west][inner sep=0.75pt]    {$\mathbf{X} :=X_{1}^{( 0)}$};
\draw (111,67.4) node [anchor=north west][inner sep=0.75pt]    {$ \begin{array}{l}
X_{1}^{( 1)}\\
\vdots \\
X_{n}^{( 1)}
\end{array}$};
\draw (250,27.4) node [anchor=north west][inner sep=0.75pt]    {$ \begin{array}{l}
X_{1}^{( L)} :=X_{1}\\
\vdots \\
\\
\vdots \\
\\
\vdots \\
X_{n^{L}}^{( L)} :=X_{n^{L}}
\end{array}$};
\draw (189.5,93.4) node [anchor=north west][inner sep=0.75pt]    {$\cdots $};

\end{tikzpicture}
\end{minipage}
\bigskip

\noindent where the variables from the last layer are exactly the random variables that will be determined by the noisy random variables
\[ X_1^{(L)}\otimes\cdots\otimes X^{(L)}_{n^L}:=X_1\otimes\cdots\otimes X_{n^L}.\]
Since the error correction is performed by layers, we can furthermore define the $l$-\emph{weight} of $X^{(l)}_j$ for every $j$ such that
\begin{equation}\begin{split}
\omega^{(L)}(X^{(L)}_j)&:= \omega(X_j)=\begin{cases}1 & \text{if } X_j\not=I\\
0 & \text{if } X_j=I\end{cases},\quad\text{for } l=L,\\
\omega^{(l)}(X^{(l)}_j)&:=\begin{cases}1 & \text{if } X^{(l)}_j\in (S^{(L-l+1)})^\bot\setminus S^{(L-l+1)}\\
0 & \text{else}\end{cases},\quad\text{for } l=0,\ldots, L-1.
\end{split}\end{equation}
Such that the $0$-th level determines if the error correction decodes the message appropriately when dealing with noise or not. For clarity, we will drop the superscripts and denote by $\textbf{X}:=X^{(0)}$ the random variable determining the success or failure of the error correction, and by $S:=S^{(L)}$ the stabilizer group of the $[[n^L,1,d^L]]$ code. Note that for each level $l$ the $\{X_1^{(l)},\ldots,X_{n^l}^{(l)}\}$ random variables are independent and identially distributed (i.i.d.). Since the error-correcting code can correct any error in $t$ qubits, we can bound the probability of failed recovery by
\begin{equation}\begin{split} \Pr(\omega(\textbf{X})=1)&=\Pr(\textbf{X}\in S^\bot\setminus S)\leq \sum_{k=t+1}^n\Pr(\sum_{j=1}^n\omega(X^{(1)}_j)=k)\\
&=\sum_{k=t+1}^n\binom{n}{k}\Pr(\omega(X^{(1)})=1)^k\left(1-\Pr(\omega(X^{(1)})=1)\right)^{n-k}\\
&\leq\binom{n}{t+1}\Pr(\omega(X^{(1)})=1)^{t+1}. \end{split}\end{equation}
Applying the above argument iteratively we can show that after $L$ levels of concatenation, the probability of failed recovery is upper bounded by 
\begin{equation}\label{eq:threshold} \Pr(\textbf{X}\in S^\bot\setminus S) \leq \binom{n}{t+1}^{-1}\left(\binom{n}{t+1}p\right)^{(t+1)^L}.\end{equation}
Note that if $p<\pth:=\binom{n}{t+1}^{-1}$, then we can make the failure probability as small as desired by increasing the number of layers. In the context of quantum computing, this upper value for the probability of single-qubit errors is known as the \emph{threshold value}, discovered by Aharonov et al.~\cite{aharonov_fault-tolerant_2008} for general error-correcting codes.

\subsection{Quantum authentication}\label{sec:authentication}

The goal of authentication is to verify the integrity of a message. In order to achieve this, the sender and receiver make use of polynomial-time keyed pairs of encoding and decoding maps, such that an adversary without knowledge of the key, who tampers with the data, will be discovered with very high probability. This quantification of tampering is a nontrivial task, which has lead to multiple works extending and rigorously defining Barnum et al.'s~\cite{barnum_authentication_2002} proposed original notion of quantum authentication. 


In the context of constructive cryptography, a quantum authentication protocol is expected to construct an \emph{authenticated quantum channel}, $\mathcal{S}$, from nothing but an insecure quantum channel and a secret key source. The goal of a secure quantum channel is to allow Alice to send $m$ qubits to Bob without Eve tampering with the data. On the one hand, they cannot stop Eve from learning that a message has been transmitted nor cutting the communication lines. Hence Eve's actions can be described as a bit $0$ when Bob gets the message, and $1$ when he does not. On the other hand, in the presence of no adversary, Eve's interface is substituted by a filter $\Diamond_E$ that models an honest behaviour, in this case always allowing Bob to receive exactly the message that Alice sent. \Cref{fig:secure_channel} is a graphical description of the channel $\mathcal{S}_\Diamond$. 
\begin{figure}
\centering
    \begin{subfigure}{0.45\textwidth}
        \centering
        \begin{tikzpicture}[x=0.75pt,y=0.75pt,yscale=-1,xscale=1]

\draw   (60.38,19.99) -- (220.88,19.99) -- (220.88,81.49) -- (60.38,81.49) -- cycle ;
\draw    (168.17,56.26) -- (168.09,110.74) ;
\draw  [dash pattern={on 4.5pt off 4.5pt}]  (111.28,50.61) -- (111.37,107.99) ;
\draw [shift={(111.38,109.99)}, rotate = 269.91] [color={rgb, 255:red, 0; green, 0; blue, 0 }  ][line width=0.75]    (10.93,-3.29) .. controls (6.95,-1.4) and (3.31,-0.3) .. (0,0) .. controls (3.31,0.3) and (6.95,1.4) .. (10.93,3.29)   ;
\draw    (40.51,50.12) -- (149.88,49.99) -- (180.38,59.99) ;
\draw    (180.34,50.55) -- (246.68,50.2) ;
\draw [shift={(249.68,50.18)}, rotate = 179.7] [fill={rgb, 255:red, 0; green, 0; blue, 0 }  ][line width=0.08]  [draw opacity=0] (8.93,-4.29) -- (0,0) -- (8.93,4.29) -- cycle    ;
\draw    (171.26,56.76) -- (171.21,107.41) -- (171.21,110.77) ;

\draw (125.23,98.62) node [anchor=north west][inner sep=0.75pt]   [align=left] {Eve};
\draw (90.9,90.05) node [anchor=north west][inner sep=0.75pt]    {$m$};
\draw (175.66,89.61) node [anchor=north west][inner sep=0.75pt]    {$0,1$};
\draw (222.88,22.99) node [anchor=north west][inner sep=0.75pt]   [align=left] {Bob};
\draw (255.51,43.2) node [anchor=north west][inner sep=0.75pt]    {$\rho ,\ \bot $};
\draw (25.2,22.57) node [anchor=north west][inner sep=0.75pt]   [align=left] {Alice};
\draw (23.5,46) node [anchor=north west][inner sep=0.75pt]    {$\rho $};

\end{tikzpicture}
        \caption{Secure authenticated quantum channel with adversary Eve.}
    \end{subfigure}\hfill
    \begin{subfigure}{0.45\textwidth}
        \centering
        \begin{tikzpicture}[x=0.75pt,y=0.75pt,yscale=-1,xscale=1,scale=0.9]

\draw   (60.38,19.99) -- (220.88,19.99) -- (220.88,81.49) -- (60.38,81.49) -- cycle ;
\draw    (168.17,56.26) -- (168.09,110.74) ;
\draw  [dash pattern={on 4.5pt off 4.5pt}]  (111.28,50.61) -- (111.37,107.99) ;
\draw [shift={(111.38,109.99)}, rotate = 269.91] [color={rgb, 255:red, 0; green, 0; blue, 0 }  ][line width=0.75]    (10.93,-3.29) .. controls (6.95,-1.4) and (3.31,-0.3) .. (0,0) .. controls (3.31,0.3) and (6.95,1.4) .. (10.93,3.29)   ;
\draw    (40.51,50.12) -- (149.88,49.99) -- (180.38,59.99) ;
\draw    (180.34,50.55) -- (246.68,50.2) ;
\draw [shift={(249.68,50.18)}, rotate = 179.7] [fill={rgb, 255:red, 0; green, 0; blue, 0 }  ][line width=0.08]  [draw opacity=0] (8.93,-4.29) -- (0,0) -- (8.93,4.29) -- cycle    ;
\draw    (171.26,56.76) -- (171.21,107.41) -- (171.21,110.77) ;
\draw   (94.5,114.3) .. controls (94.5,111.92) and (96.42,110) .. (98.8,110) -- (181.48,110) .. controls (183.86,110) and (185.78,111.92) .. (185.78,114.3) -- (185.78,127.2) .. controls (185.78,129.57) and (183.86,131.5) .. (181.48,131.5) -- (98.8,131.5) .. controls (96.42,131.5) and (94.5,129.57) .. (94.5,127.2) -- cycle ;

\draw (90.9,90.05) node [anchor=north west][inner sep=0.75pt]    {$m$};
\draw (175.66,89.61) node [anchor=north west][inner sep=0.75pt]    {$0$};
\draw (222.88,22.99) node [anchor=north west][inner sep=0.75pt]   [align=left] {Bob};
\draw (255.51,43.2) node [anchor=north west][inner sep=0.75pt]    {$\rho ,\ \bot $};
\draw (22.2,22.57) node [anchor=north west][inner sep=0.75pt]   [align=left] {Alice};
\draw (23.5,46) node [anchor=north west][inner sep=0.75pt]    {$\rho $};
\draw (189.5,122.4) node [anchor=north west][inner sep=0.75pt]    {$\diamondsuit $};

\end{tikzpicture}
        \caption{Secure authenticated quantum channel with no adversary present.}\label{fig:secure_noeve}
    \end{subfigure}
\caption{Characterization of an authenticated quantum channel $\mathcal{S}_\Diamond$.}\label{fig:secure_channel}
\end{figure}
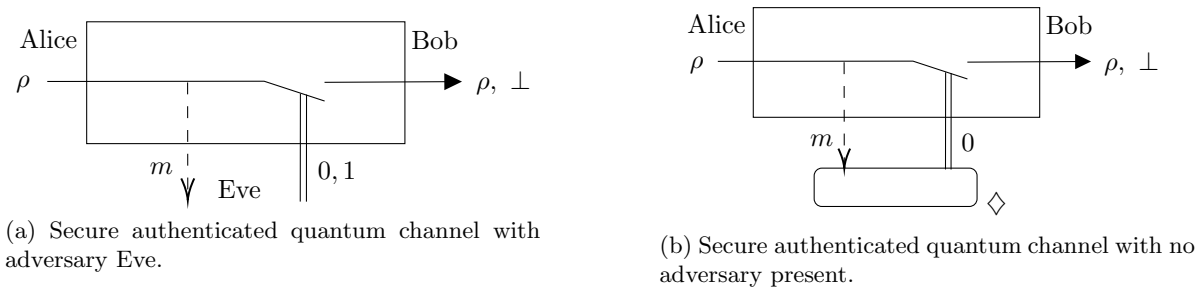
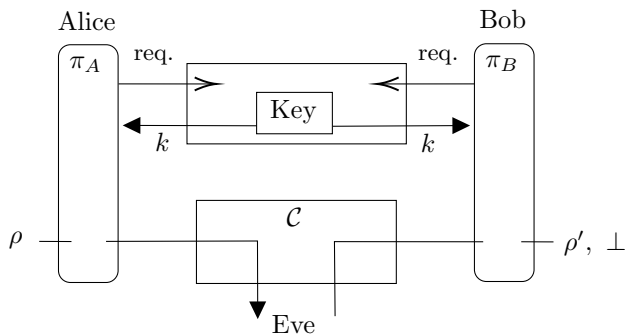
\begin{figure}
\centering
    \centering
    \begin{tikzpicture}[x=0.75pt,y=0.75pt,yscale=-1,xscale=1]

\draw   (30.38,36.21) .. controls (30.38,32.89) and (33.07,30.2) .. (36.39,30.2) -- (54.43,30.2) .. controls (57.76,30.2) and (60.45,32.89) .. (60.45,36.21) -- (60.45,144.19) .. controls (60.45,147.51) and (57.76,150.2) .. (54.43,150.2) -- (36.39,150.2) .. controls (33.07,150.2) and (30.38,147.51) .. (30.38,144.19) -- cycle ;
\draw   (240.16,35.71) .. controls (240.16,32.39) and (242.85,29.7) .. (246.17,29.7) -- (264.22,29.7) .. controls (267.54,29.7) and (270.23,32.39) .. (270.23,35.71) -- (270.23,143.69) .. controls (270.23,147.01) and (267.54,149.7) .. (264.22,149.7) -- (246.17,149.7) .. controls (242.85,149.7) and (240.16,147.01) .. (240.16,143.69) -- cycle ;
\draw   (95.19,39.7) -- (205.81,39.7) -- (205.81,80.24) -- (95.19,80.24) -- cycle ;
\draw    (168.64,71.35) -- (234.98,71) ;
\draw [shift={(237.98,70.98)}, rotate = 179.7] [fill={rgb, 255:red, 0; green, 0; blue, 0 }  ][line width=0.08]  [draw opacity=0] (8.93,-4.29) -- (0,0) -- (8.93,4.29) -- cycle    ;
\draw   (130.3,54.25) -- (168.51,54.25) -- (168.51,75.14) -- (130.3,75.14) -- cycle ;

\draw    (130.44,71.1) -- (65.76,70.92) ;
\draw [shift={(62.76,70.91)}, rotate = 0.16] [fill={rgb, 255:red, 0; green, 0; blue, 0 }  ][line width=0.08]  [draw opacity=0] (8.93,-4.29) -- (0,0) -- (8.93,4.29) -- cycle    ;

\draw    (240.34,49.96) -- (193.01,49.99) ;
\draw [shift={(191.01,49.99)}, rotate = 359.96] [color={rgb, 255:red, 0; green, 0; blue, 0 }  ][line width=0.75]    (10.93,-3.29) .. controls (6.95,-1.4) and (3.31,-0.3) .. (0,0) .. controls (3.31,0.3) and (6.95,1.4) .. (10.93,3.29)   ;
\draw    (60.67,49.96) -- (108.02,49.99) ;
\draw [shift={(110.02,49.99)}, rotate = 180.04] [color={rgb, 255:red, 0; green, 0; blue, 0 }  ][line width=0.75]    (10.93,-3.29) .. controls (6.95,-1.4) and (3.31,-0.3) .. (0,0) .. controls (3.31,0.3) and (6.95,1.4) .. (10.93,3.29)   ;
\draw   (99.96,108.77) -- (200.76,108.77) -- (200.76,150.7) -- (99.96,150.7) -- cycle ;
\draw    (54.37,129.46) -- (130.74,129.26) -- (130.86,165.7) ;
\draw [shift={(130.88,168.7)}, rotate = 269.8] [fill={rgb, 255:red, 0; green, 0; blue, 0 }  ][line width=0.08]  [draw opacity=0] (8.93,-4.29) -- (0,0) -- (8.93,4.29) -- cycle    ;
\draw    (20.67,129.46) -- (36.93,129.53) ;
\draw    (263.77,129.64) -- (276.37,129.7) -- (280.03,129.71) ;
\draw    (244.58,129.65) -- (169.56,129.58) -- (169.88,167.2) ;

\draw (241.38,10.99) node [anchor=north west][inner sep=0.75pt]   [align=left] {Bob};
\draw (28.7,12.07) node [anchor=north west][inner sep=0.75pt]   [align=left] {Alice};
\draw (135.9,56.65) node [anchor=north west][inner sep=0.75pt]   [align=left] {Key};
\draw (67.06,31.4) node [anchor=north west][inner sep=0.75pt]   [align=left] {{\small req.}};
\draw (210.56,31.5) node [anchor=north west][inner sep=0.75pt]   [align=left] {{\small req.}};
\draw (34.75,34.9) node [anchor=north west][inner sep=0.75pt]    {$\pi _{A}$};
\draw (244.55,34.4) node [anchor=north west][inner sep=0.75pt]    {$\pi _{B}$};
\draw (144.27,111.4) node [anchor=north west][inner sep=0.75pt]    {$\mathcal{C}$};
\draw (284.05,122.7) node [anchor=north west][inner sep=0.75pt]    {$\rho' ,\ \bot $};
\draw (4,123.33) node [anchor=north west][inner sep=0.75pt]    {$\rho $};
\draw (136.73,165.12) node [anchor=north west][inner sep=0.75pt]   [align=left] {Eve};
\draw (77.85,73.4) node [anchor=north west][inner sep=0.75pt]    {$k$};
\draw (211.85,73.8) node [anchor=north west][inner sep=0.75pt]    {$k$};

\end{tikzpicture}
    \caption{The real system for quantum message authentication.}
\end{figure}

In order to construct the filtered resource $\mathcal{S}_\Diamond$, quantum authentication protocols will use a shared secret key $\mathcal{K}$ and an insecure quantum channel $\mathcal{C}_\#$, here the filter $\Diamond_E$ represents an honest behaviour of the adversary, and the filter $\#_E$ is a noisy channel. After receiving a message $\rho$, the protocol $\pi_A$ encrypts it with the key $k$ received from $\mathcal{K}$ and sends the message to the insecure quantum channel $\mathcal{C}$. The protocol $\pi_B$ upon receiving a message checks its validity with the shared secret key $k$, and outputs either $\rho'$ or an error message $\bot$. In absence of an adversary, we substitute Eve's interface by a noise filter $\#_E$. Note that for our purposes we are not considering key resources, which greatly simplifies Portmann's descriptions~\cite[Section 3]{portmann_quantum_2017}.


A generic way of constructing authentication codes was given by Barnum et al.~\cite{barnum_authentication_2002} using purity-testing codes. In these schemes, the message is first encrypted with a quantum one-time pad using the shared secret key, and then encoded using a $[[n,1,d]]$ purity-testing error-correcting code~$\{V_k\}_{k\in\mathcal{K}}$ and a random syndrome, see \Cref{fig:construction_purity}. These schemes are also called `encrypt-then-encode' schemes, and it is not difficult to prove that the reverse order `encode-then-encrypt' provides equivalent security results~\cite{portmann_quantum_2017}. Note that the schemes differ in the amount of secret key needed, but since we are not going to focus on key consumption here, we refer the interested reader to the aforementioned article.

\begin{figure}
\begin{center}
    \begin{tabular}{ |c p{14cm}| }
    \hline
    \multicolumn{2}{|l|}{\textbf{Construction}: quantum authentication `encrypt-then-encode' scheme.} \\
    \hline
    \hline
    \multicolumn{2}{|l|}{\textbf{Encoding:}} \\
    1. & Alice and Bob obtain uniform keys $k$ (to choose a purity-testing code), $l$ (for the encryption) and $s$ (the error syndrome) from the key resource.\\
    2. & Alice encrypts the message $\rho_A$ with a quantum one-time pad using the key $l$. She appends a $n$-qubit state $\dyad{s}_S$, and encodes everything with a purity testing code, obtaining
    \[ \sigma_{AS}=V_k(P_l\rho_A P_l\otimes\dyad{s}_S)V_k^\dag. \]\\
    3. & Alice sends $\sigma_{AS}$ to Bob through the insecure quantum channel.\\
    \hline
    \multicolumn{2}{|l|}{\textbf{Decoding:}} \\
    1. & Bob receives $\hat{\sigma}_{AS}$ and measures the syndrome $S$ of the error-correcting code $V_k$ in the computational basis. If the syndrome $\hat{s}$ coincides with $s$ he accepts, else he aborts the protocol.\\
    2. & If Bob accepts the protocol, applied the decoding unitary $V_k$ and decrypts the data using the key $l$.\\
    \hline
    \end{tabular}
\end{center}
\caption{QMA from purity-testing codes.}\label{fig:construction_purity}
\end{figure}


We previously defined a quantum message authentication system as a protocol that constructs an authenticated quantum channel $\mathcal{S}_\Diamond$ as in \Cref{fig:secure_channel}, from some shared secret key $\mathcal{K}$ and an insecure quantum channel $\mathcal{C}_\#$, where the filter introduces noise. Portmann showed that the scheme from~\Cref{fig:construction_purity} based on purity-testing codes provides quantum authentication protocols, given that the filter is noiseless, denoted $\square_E$. 

\begin{theorem}[\cite{portmann_quantum_2017}, Lemma~D.1]\label{thm:sec_purity}
Given a $\delta$-purity testing protocol $[[n,1,d]]$, let $\pi^{\text{auth}}_{AB}=(\pi_A,\pi_B)$ denote the converter corresponding to Alice and Bob's protocols from \Cref{fig:construction_purity}. Then $\pi^{\text{auth}}_{AB}$  constructs an authenticated quantum channel $\mathcal{S}_\Diamond$, given an insecure noiseless quantum channel $\mathcal{C}_\square$ and a secret shared key $\mathcal{K}$ within $(0,\delta^{\text{auth}})$, where $\delta^{\text{auth}}=\max\{\delta,2^{-(n-1)}\}$. That is,
\[ \mathcal{C}_\square||\mathcal{K}\xrightarrow{\pi^{\text{auth}}_{AB}, (0,\delta^{\text{auth}})}\mathcal{S}_\Diamond. \]
\end{theorem}

\subsection{Noisy channels}\label{sec:noise}

The quantum authentication protocols described in~\Cref{sec:authentication} reject as soon as an error is present, but realistic channels are naturally noisy. The first attempt to solve the problem by wrapping the authentication protocol with error correction was proposed in~\cite{hayden_universal_2016}, and later made explicit in~\cite{portmann_quantum_2017}, where security of the composed protocol also proven. In other words, a composition of an authentication protocol and an error-correcting protocol -- note that a composable security notion is essential here -- can construct authenticated quantum channels from nothing but noisy insecure quantum channels and a shared secret key.

Given a noisy quantum channel between Alice $A$ and Bob $B$, where the \emph{noise} is represented by a quantum operation $\mathcal{F}_{A\to B}$, we say that there exists an error-correction protocol $\pi_{AB}^{\text{ecc}}$, defined by an encoding map $\mathcal{E}_A$ and a decoding map $\mathcal{D}_B$, correcting the errors induced by $\mathcal{F}_{A\to B}$ within $\eps^{\text{ecc}}$, if 
\[ \frac{1}{2}\left\|\mathcal{D}_B\circ\mathcal{F}_{A\to B}\circ\mathcal{E}_A-I_{A\to B}\right\|_\diamond\leq\eps^{\text{ecc}}. \]

We can rewrite the above statement in the abstract cryptography language.
\begin{lemma}[\cite{portmann_quantum_2017}, Lemma~4.2]\label{lem:error_cor}
Let $\#_E$ be a filter introducing the noise given by the quantum operation~$\mathcal{F}$, and let $\square_E$ be a noiseless filter. If there exists an error correction protocol $\pi_{AB}^{\text{ecc}}=(\pi^{\text{ecc}}_A,\pi^{\text{ecc}}_B)$ that corrects the errors induced by $\mathcal{F}$ within $\eps^{ecc}$, then $\pi_{AB}^{\text{ecc}}$ constructs a noiseless channel $\mathcal{C}_{\square}$, from a noisy channel $\hat{\mathcal{C}}_{\#}$ within $(\eps^{\text{ecc}},0)$. That is,
\[ \hat{\mathcal{C}}_{\#}\xrightarrow{\pi^{\text{ecc}}_{AB},(\eps^{\text{ecc}},0)}\mathcal{C}_\square. \]
\end{lemma}


It is now clear what the relevance of splitting the cryptographic security definition in terms of correctness and security is, it allows us to prove that a $\delta$-secure authentication scheme wrapped in an $\eps$-correct error-correcting code constructs an $(\eps,\delta)$-secure authenticated quantum channel, instead of an $(\eps+\delta)$-secure one as the original composition theorem would provide.
\begin{theorem}
Let $\pi_{AB}^{\text{auth}}$ be a $(0,\delta^{\text{auth}})$-secure `encode-then-encrypt' authentication protocol from~\Cref{fig:construction_purity}, and let $\pi_{AB}^{\text{ecc}}$ be an $(\eps^{ecc},0)$-secure error-correcting protocol dealing with $\#_E$ noise from~\Cref{lem:error_cor}. The composition of these protocols constructs an authenticated quantum channel $\mathcal{S}_\Diamond$, from an insecure noisy quantum channel $\hat{\mathcal{C}}_\#$, and shared secret key $\mathcal{K}$ within $(\eps^{\text{ecc}},\delta^{\text{auth}})$. That is,
\begin{equation}\label{eq:composition} \hat{\mathcal{C}}_\#||\mathcal{K}\xrightarrow{\pi_{AB}^{\text{auth}}\pi_{AB}^{\text{ecc}}, (\eps^{\text{ecc}},\delta^{\text{auth}})} \mathcal{S}_\Diamond.
\end{equation}
\begin{proof}
Direct consequence of~\Cref{thm:sec_purity,thm:composition}, and~\Cref{lem:error_cor}.
\end{proof}
\end{theorem}

\begin{figure}
\centering
    \centering
    \begin{tikzpicture}[x=0.75pt,y=0.75pt,yscale=-1,xscale=1]

\draw   (36.71,39.97) .. controls (36.71,34.94) and (40.78,30.87) .. (45.81,30.87) -- (73.12,30.87) .. controls (78.15,30.87) and (82.23,34.94) .. (82.23,39.97) -- (82.23,141.76) .. controls (82.23,146.79) and (78.15,150.87) .. (73.12,150.87) -- (45.81,150.87) .. controls (40.78,150.87) and (36.71,146.79) .. (36.71,141.76) -- cycle ;
\draw   (334.66,39.31) .. controls (334.66,34.28) and (338.74,30.2) .. (343.77,30.2) -- (371.1,30.2) .. controls (376.13,30.2) and (380.21,34.28) .. (380.21,39.31) -- (380.21,141.09) .. controls (380.21,146.12) and (376.13,150.2) .. (371.1,150.2) -- (343.77,150.2) .. controls (338.74,150.2) and (334.66,146.12) .. (334.66,141.09) -- cycle ;
\draw   (153.36,39.2) -- (263.97,39.2) -- (263.97,79.74) -- (153.36,79.74) -- cycle ;
\draw    (226.81,70.85) -- (325.04,70.88) ;
\draw [shift={(328.04,70.88)}, rotate = 180.02] [fill={rgb, 255:red, 0; green, 0; blue, 0 }  ][line width=0.08]  [draw opacity=0] (8.93,-4.29) -- (0,0) -- (8.93,4.29) -- cycle    ;
\draw   (188.46,53.75) -- (226.67,53.75) -- (226.67,74.64) -- (188.46,74.64) -- cycle ;

\draw    (188.61,70.6) -- (92,70.6) ;
\draw [shift={(89,70.6)}, rotate = 360] [fill={rgb, 255:red, 0; green, 0; blue, 0 }  ][line width=0.08]  [draw opacity=0] (8.93,-4.29) -- (0,0) -- (8.93,4.29) -- cycle    ;
\draw    (334.4,49.45) -- (251.18,49.49) ;
\draw [shift={(249.18,49.49)}, rotate = 359.97] [color={rgb, 255:red, 0; green, 0; blue, 0 }  ][line width=0.75]    (10.93,-3.29) .. controls (6.95,-1.4) and (3.31,-0.3) .. (0,0) .. controls (3.31,0.3) and (6.95,1.4) .. (10.93,3.29)   ;
\draw    (82.17,49.45) -- (166.18,49.49) ;
\draw [shift={(168.18,49.49)}, rotate = 180.03] [color={rgb, 255:red, 0; green, 0; blue, 0 }  ][line width=0.75]    (10.93,-3.29) .. controls (6.95,-1.4) and (3.31,-0.3) .. (0,0) .. controls (3.31,0.3) and (6.95,1.4) .. (10.93,3.29)   ;
\draw   (158.12,108.27) -- (258.92,108.27) -- (258.92,150.2) -- (158.12,150.2) -- cycle ;
\draw    (139.1,128.67) -- (188.9,128.76) -- (189.03,165.2) ;
\draw [shift={(189.04,168.2)}, rotate = 269.8] [fill={rgb, 255:red, 0; green, 0; blue, 0 }  ][line width=0.08]  [draw opacity=0] (8.93,-4.29) -- (0,0) -- (8.93,4.29) -- cycle    ;
\draw    (278.1,129.01) -- (227.73,129.08) -- (228.04,166.7) ;
\draw    (20.67,129.46) -- (36.93,129.53) ;
\draw    (374.21,128.94) -- (386.81,129) -- (390.47,129.01) ;
\draw   (98.77,117.14) .. controls (98.77,112.83) and (102.26,109.34) .. (106.57,109.34) -- (131.3,109.34) .. controls (135.61,109.34) and (139.1,112.83) .. (139.1,117.14) -- (139.1,140.54) .. controls (139.1,144.85) and (135.61,148.34) .. (131.3,148.34) -- (106.57,148.34) .. controls (102.26,148.34) and (98.77,144.85) .. (98.77,140.54) -- cycle ;
\draw   (278.1,117.14) .. controls (278.1,112.83) and (281.59,109.34) .. (285.9,109.34) -- (310.63,109.34) .. controls (314.94,109.34) and (318.43,112.83) .. (318.43,117.14) -- (318.43,140.54) .. controls (318.43,144.85) and (314.94,148.34) .. (310.63,148.34) -- (285.9,148.34) .. controls (281.59,148.34) and (278.1,144.85) .. (278.1,140.54) -- cycle ;
\draw    (318.32,128.81) -- (330.92,128.86) -- (334.58,128.88) ;
\draw    (82.32,128.81) -- (94.92,128.86) -- (98.58,128.88) ;
\draw  [color={rgb, 255:red, 74; green, 144; blue, 226 }  ,draw opacity=1 ] (30.65,47.31) .. controls (30.65,34.57) and (40.98,24.24) .. (53.72,24.24) -- (122.92,24.24) .. controls (135.66,24.24) and (145.98,34.57) .. (145.98,47.31) -- (145.98,136.51) .. controls (145.98,149.25) and (135.66,159.58) .. (122.92,159.58) -- (53.72,159.58) .. controls (40.98,159.58) and (30.65,149.25) .. (30.65,136.51) -- cycle ;
\draw  [color={rgb, 255:red, 74; green, 144; blue, 226 }  ,draw opacity=1 ] (271.32,47.31) .. controls (271.32,34.57) and (281.64,24.24) .. (294.38,24.24) -- (363.58,24.24) .. controls (376.32,24.24) and (386.65,34.57) .. (386.65,47.31) -- (386.65,136.51) .. controls (386.65,149.25) and (376.32,159.58) .. (363.58,159.58) -- (294.38,159.58) .. controls (281.64,159.58) and (271.32,149.25) .. (271.32,136.51) -- cycle ;

\draw (311.21,5.49) node [anchor=north west][inner sep=0.75pt]   [align=left] {Bob};
\draw (72.03,5.4) node [anchor=north west][inner sep=0.75pt]   [align=left] {Alice};
\draw (194.06,56.15) node [anchor=north west][inner sep=0.75pt]   [align=left] {Key};
\draw (42.75,34.5) node [anchor=north west][inner sep=0.75pt]    {$\pi _{A}^{auth}$};
\draw (339.05,34.9) node [anchor=north west][inner sep=0.75pt]    {$\pi _{B}^{auth}$};
\draw (394.15,122) node [anchor=north west][inner sep=0.75pt]    {$\rho ,\ \bot $};
\draw (4,123.33) node [anchor=north west][inner sep=0.75pt]    {$\rho $};
\draw (101.9,30.23) node [anchor=north west][inner sep=0.75pt]   [align=left] {{\small req.}};
\draw (289.4,31) node [anchor=north west][inner sep=0.75pt]   [align=left] {{\small req.}};
\draw (202.44,110.9) node [anchor=north west][inner sep=0.75pt]    {$\mathcal{C}$};
\draw (194.9,164.62) node [anchor=north west][inner sep=0.75pt]   [align=left] {Eve};
\draw (99.47,74) node [anchor=north west][inner sep=0.75pt]    {$k$};
\draw (306.02,73.3) node [anchor=north west][inner sep=0.75pt]    {$k$};
\draw (104.27,115.56) node [anchor=north west][inner sep=0.75pt]    {$\pi _{A}^{ecc}$};
\draw (284.94,116.24) node [anchor=north west][inner sep=0.75pt]    {$\pi _{B}^{ecc}$};
\draw (79,162) node [anchor=north west][inner sep=0.75pt]  [color={rgb, 255:red, 74; green, 144; blue, 226 }  ,opacity=1 ]  {$\pi _{A}$};
\draw (319.12,162) node [anchor=north west][inner sep=0.75pt]  [color={rgb, 255:red, 74; green, 144; blue, 226 }  ,opacity=1 ]  {$\pi _{B}$};

\end{tikzpicture}
    \caption{The real system for the composed error correction and authentication protocol.}
\end{figure}
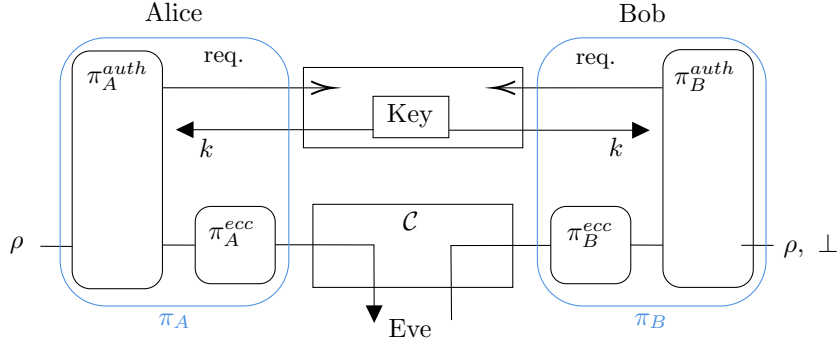


In order to be able to compare explicit schemes, we will restrict to a basic type of noise, typically used in error correction literature, i.i.d.\ Pauli noise. We will assume that when qubits are sent through a noisy channel, they independently undergo a $X$, $Y$ or $Z$ Pauli error with probabilities $p_X$, $p_Y$ and $p_Z$ respectively. This model is interesting not only because it models many interesting real situations, but also because the Pauli operators are a basis of single-qubit operations, and thus protection against i.i.d.\ Pauli noise for a single qubit implies protection against any single-qubit error. Moreover, in the `encode-then-encrypt' authentication scheme, the one-time-pad encryption and the Pauli twirl make any attack become a Pauli attack. This means that for the security proof it is enough to prove security against Pauli attacks~\cite{broadbent_quantum_2013}, which is also the reason why the underlying error-correction codes are required to be purity-testing codes.

Let $\mathcal{F}_n$ denote a quantum noise channel acting on $n$-qubits, which we can write in terms of the basis elements of single-qubit operations
\[ \mathcal{F}(\rho)=(1-p_{XYZ})\rho+p_XX\rho X^\dag+p_Y Y\rho Y^\dag + p_Z Z\rho Z^\dag,\]
where $p_{XYZ}= p_X+p_Y +p_Z$. This channel leaves the state untouched with probability $1-p_{XYZ}$ and each Pauli operation is applied with probability $p_X$, $p_Y$ and $p_Z$ respectively. As mentioned earlier, we will assume that the noise is local, i.e.\ acts independently on qubits of $n$-qubit systems. Therefore, we can write the noise channel acting on a $n$-qubit register as
\begin{equation}\label{eq:gen_noise}
\mathcal{F}_n(\rho)=\mathcal{F}^{\otimes n}(\rho)=\sum_{\substack{k_1+k_2+k_3+k_4=n\\ k_1,k_2,k_3,k_4\geq0}}\binom{n}{k_1,k_2,k_3,k_4}\prod_{j\in\{I,X,Y,Z\}}p_j^{k_j}\sigma_j^{k_j}\rho(\sigma^\dag_j)^{k_j}.
\end{equation}

The depolarizing channel, the most commonly used noise model in error-correction literature~\cite{terhal_quantum_2015}, is of this type. When a qubit goes through the depolarizing channel, the channel erases the qubit and substitutes it by a completely mixed state $I/2$ with probability $p$, and leaves the qubit untouched with probability $1-p$. In notation from \cref{eq:gen_noise}, this is the same as saying that with probability $1-p$ the qubit is being left untouched, and each Pauli operation will be applied with probability $p/3$. Therefore, the depolarizing noise acting on $n$-qubits can be written as
\begin{equation}\label{eq:depolarizing}\begin{split}
\mathcal{F}_n(\rho)&= \sum_{\substack{k_1+\cdots+k_4=n\\ k_1,\ldots,k_4\geq0}}\binom{n}{k_1,\ldots,k_4}(1-p)^{k_1}\left(\frac{p}{3}\right)^{n-k_1}\prod_{j\in\{X,Y,Z\}}\sigma_j^{k_j}\rho(\sigma^\dag_j)^{k_j},\\
& =(1-p)^n+(1-p)^{n-1}\frac{p}{3}\sum_{k=1}^n\sum_{j\in\{I,X,Y,Z\}}\sigma_j^k\rho(\sigma_j^\dag)^k+\cdots.\end{split}\end{equation}

\section{Explicit composed protocols}

Note that when we construct a noiseless channel $\mathcal{C}$ from a noisy channel $\hat{\mathcal{C}}_\#$ in \Cref{lem:error_cor} the dimension of the two channels differs, because the error-correcting codes encode logical qubits in redundant physical qubits for protection, and the same is true for authentication codes. Although we can make the failure probability of an error-correcting code arbitrarily low when undergoing a depolarizing channel, see \Cref{sec:error}, there is a trade-off in the amount of qubits required. Taking the amount of qubits necessary as a parameter for efficiency, in this section we analyze the cost-effectiveness of two well-known authentication schemes: the trap scheme and the Clifford scheme.

\subsection{Trap scheme}


The trap authentication scheme is an example of an `encode-then-encrypt' scheme, see \Cref{fig:construction_purity}, introduced by Broadbent, Gutoski, and Stebila~\cite{broadbent_quantum_2013}.
A very interesting property of this authentication scheme is its natural interaction with computation; it is possible to perform some quantum gates `transversally' on qubits of the ciphertext, which results in a valid authentication of a new ciphertext (with an updated key).
This property enabled the trap code to be a crucial ingredient in various results within quantum cryptography, such as the construction of quantum one-time programs~\cite{broadbent_quantum_2013}, a scheme for quantum zero-knowledge proofs for QMA~\cite{broadbent_zero-knowledge_2016}, and verifiable homomorphic encryption~\cite{alagic_quantum_2017}. Also see the security proof of the trap code via an efficient simulator~\cite{broadbent_efficient_2016}, and an extended version of the trap code which supports key recycling and ciphertext authentication~\cite{dulek_quantum_2018} for more context of this code.

The trap code is constructed as follows. Given a fixed $[[n,1,d]]$ error-correcting code, the scheme constructs a set of purity testing codes by appending $2n$ `traps' to the data qubits ($n$ computational-basis traps in the $\dyad{0}$ state and $n$ Hadamard-basis traps in the $\dyad{+}$ state); the resulting $3n$-qubit register is permuted in a random fashion attending to a secret shared key. When decoding, first the inverse permutation, according to the secret key, is applied. Finally, the data registers are decoded according to the fixed error-correcting code and the traps are measured in the computational and Hadamard bases respectively. Here we consider the underlying code as error \emph{correcting}, for the sake of fair comparison with later codes, but there is also an error \emph{detection} variant of the trap code~\cite{broadbent_quantum_2013}. Since the trap code is purity testing, the scheme described in~\Cref{fig:trap} gives rise to a secure authentication scheme by~\Cref{thm:sec_purity}.

\begin{figure}[h]
\begin{center}
    \begin{tabular}{ |c p{14cm}| }
    \hline
    \multicolumn{2}{|l|}{\textbf{Protocol 1}: trap authentication scheme, ``encode-then-encrypt'' form.} \\
    \hline
    \hline
    \multicolumn{2}{|l|}{\textbf{Encoding:}} \\
    1. & Alice and Bob agree on a $[[n,1,d]]$ quantum error-correcting code.\\
    2. & Alice and Bob obtain uniform keys $k$ (for the permutation) and $l$ (for the encryption) from the key resource.\\
    3. & Alice encodes the message $\rho_A$ with the agreed error-correcting code. Appends a $2n$ computational basis states $\dyad{0}^{\otimes 2n}$ and applies a Hadamard gate to the last $n$ qubits (so they are in the Hadamard-basis state $\dyad{+}$). She applies a permutation to all the qubit registers according to the secret key $k$.\\
    4. & Finally, she encrypts the message with a quantum one-time pad using the key $l$, obtaining thus
    \[ \sigma_{AS}=P_l\pi_k\left(\text{Enc}(\rho_A)\otimes\dyad{0}^{\otimes n}\otimes\dyad{+}^{\otimes n}\right)\pi^\dag_kP_l. \]\\
    5. & Alice sends $\sigma_{AS}$ to Bob through the insecure quantum channel.\\
    \hline
    \multicolumn{2}{|l|}{\textbf{Decoding:}} \\
    1. & Bob receives $\hat{\sigma}_{AS}$ and decrypts the data using $l$. Then he applies the inverse permutation according to $k$ and measures the last $2n$ registers in the computational and Hadamard bases respectively. If the measurement results in $\dyad{0}^n\otimes\dyad{+}^n$, he accepts. Else, he aborts.\\
    2. & If Bob accepts, he decodes the data register according to the agreed error-correcting code.\\
    \hline
    \end{tabular}
\end{center}
\caption{Trap authentication scheme.}\label{fig:trap}
\end{figure}

The idea of the trap code is to use the traps as a measure of the weight of the Pauli operators applied to the encoded message. Since the error correction code can deal well with low-weight errors, and the traps very efficiently find high-weight ones, the probability of non-trivial errors being undetected by the code is low, i.e.\ it is a purity testing code. The previous analysis of the purity testing parameter by Broadbent and Wainewright~\cite{broadbent_efficient_2016} does not exploit the full power of the underlying error-correcting code since these codes indeed perfectly correct sub-linear weight Pauli errors, but also correct with high probability linear weight Pauli errors. However, the security parameter is still as tight as the one for the threshold code, leading to a fair comparison. For a detailed analysis we refer the reader to~\Cref{app:trap}.

\begin{lemma}[\cite{broadbent_efficient_2016}]
The trap code with inner error-correcting code $[[n,1,d]]$ is $(1/3)^{\frac{d+1}{2}}$-purity testing.
\end{lemma}

Given a purity testing code, the discussion in \Cref{sec:authentication,sec:noise} ensures us that if there exists an error-correcting protocol correcting the errors induced by the noisy channel $\mathcal{F}$ within $\eps^{ecc}$, then the composed protocol will be $\eps^{ecc}$-correct and $(1/3)^{\frac{d+1}{2}}$-secure. The following theorem rephrases this statement in terms of the efficiency of the composed protocol, where for the sake of comparison analysis we take the depolarizing noise, see \cref{eq:depolarizing}, and a concatenated error-correcting code, see \Cref{sec:error}.
\begin{prop}\label{prop:trap}
Let $\#_E$ be a filter introducing the noise given by the depolarizing channel with channel error $p<\pth$. If the trap authentication scheme $\pi^{\text{trap}}$ with an $[n^M_1,1,d^M_1]$ inner code is composed with a $[[n_2^L,1,d_2^L]]$ concatenated error-correcting code $\pi^{\text{ecc}}$, then to obtain $\eps$-correctness and $\delta$-security, i.e.
\[ \hat{\mathcal{C}}_\#||\mathcal{K}^\mu\xrightarrow{\pi^{\text{trap}}\pi^{\text{ecc}},(\eps,\delta)}\mathcal{S}^m_\Diamond, \]
it is sufficient for the amount of qubits to grow as
\[ O\left(\log(1/\eps)^{C_2}\log(1/\delta)^{C_1}\right). \]
Here the constants $C_1(t_1,n_1):=\log_{2t_1+1}(n_1)$ and $C_2(t_2,n_2):=\log_{t_2+1}(n_2)$ depend only on the properties of the error-correcting codes chosen for the concatenation.

\begin{proof}
    Recall that in \cref{eq:composition} the correctness and security of the protocol only depend on the error-correcting capabilities of the outer error-correcting and the authenticating purity-testing codes respectively. However, since data qubits are first encoded with a fixed inner error-correcting code (which also uniquely determines the size of the purity testing code) and later encoded again in the outer error-correcting code, the total size of the protocol depends on both parameters.
    
    It is most natural to start with the size of the inner code, which determines security. Note that the amount of errors that an error-correcting code can correct is a function of the distance, which is at the same time a function of the total amount of errors that each layer of concatenation can correct, i.e. 
    \[ \frac{d_1^M+1}{2}=\frac{(2t_1+1)^M+1}{2}. \]
    Then $\delta$-security is obtained whenever
    \begin{equation}\label{eq:sec_trap}\begin{split} (1/3)^{\frac{d+1}{2}}\leq\delta &\Leftrightarrow \left(\frac{(2t_1+1)^M+1}{2}\right)\log(1/3)\leq\log(\delta) \Leftrightarrow (2t_1+1)^M\geq\frac{2\log(1/\delta)}{\log(3)}-1\\
    & \Leftrightarrow n_1^M\gtrsim \log(1/\delta)^{\log_{2t_1+1}(n_1)}. \end{split}\end{equation}
    For the final step we used the well known property of the logarithm $a^b=(x^b)^{\log_x(a)}$.
    
    On the other hand, an increase of levels of concatenation of the outer code is what improves the correctness of the composed protocol. As mentioned in \Cref{sec:noise}, the error of an error-correcting code is the probability of failed recovery of encoded data after going through a noisy channel. In the case of the depolarizing channel, \cref{eq:depolarizing}, by the structure of concatenated codes, whenever the error of the depolarizing channel $p$ is smaller than the threshold
    \[ p< \pth:=\binom{n_2}{t_2+1}^{-1}, \]
    we can make failure probability as small as desired by increasing the levels of concatenation. However, now the amount of qubits that have to be decoded depends on the inner code as well, see~\Cref{fig:composed}. 
    Let us denote by $X_1,\ldots, X_{3n_1^Mn_2^L}$ the random variables such that $X_j\in\{X,Y,Z\}$ with probability $p$ and $X_j=I$ with probability $1-p$, and by $\textbf{X}_1,\cdots,X_{3n_1^M}$ the random variables determining the success or failure of the outer error correction as in~\Cref{sec:error}. Moreover, $\textbf{X}^{(0)}$ will determine the success or failure of the error correction on the data, while the random variables $\text{X}_{n_1^M+1},\dots,\textbf{X}_{3n_1^M}$ refer to the $2n_1^M$-traps that are appended after the inner error-correction encoding.
    \begin{figure}
        \centering
        \begin{tikzpicture}[x=0.75pt,y=0.75pt,yscale=-1,xscale=1]

\draw    (221.5,49.65) -- (256.09,67.37) ;
\draw [shift={(257.88,68.28)}, rotate = 207.11] [color={rgb, 255:red, 0; green, 0; blue, 0 }  ][line width=0.75]    (10.93,-3.29) .. controls (6.95,-1.4) and (3.31,-0.3) .. (0,0) .. controls (3.31,0.3) and (6.95,1.4) .. (10.93,3.29)   ;
\draw    (221.5,49.65) -- (256.13,30.25) ;
\draw [shift={(257.88,29.28)}, rotate = 150.74] [color={rgb, 255:red, 0; green, 0; blue, 0 }  ][line width=0.75]    (10.93,-3.29) .. controls (6.95,-1.4) and (3.31,-0.3) .. (0,0) .. controls (3.31,0.3) and (6.95,1.4) .. (10.93,3.29)   ;
\draw    (221.5,49.65) -- (255.88,49.3) ;
\draw [shift={(257.88,49.28)}, rotate = 179.41] [color={rgb, 255:red, 0; green, 0; blue, 0 }  ][line width=0.75]    (10.93,-3.29) .. controls (6.95,-1.4) and (3.31,-0.3) .. (0,0) .. controls (3.31,0.3) and (6.95,1.4) .. (10.93,3.29)   ;

\draw    (272.81,59.07) ;
\draw [shift={(275.62,59.91)}, rotate = 196.66] [fill={rgb, 255:red, 0; green, 0; blue, 0 }  ][line width=0.08]  [draw opacity=0] (8.93,-4.29) -- (0,0) -- (8.93,4.29) -- cycle    ;
\draw  [draw opacity=0] (272.81,59.07) .. controls (272.81,59.07) and (272.81,59.07) .. (272.81,59.07) .. controls (267.28,59.07) and (262.8,54.82) .. (262.8,49.58) .. controls (262.8,44.34) and (267.28,40.09) .. (272.81,40.09) .. controls (278.33,40.09) and (282.82,44.34) .. (282.82,49.58) .. controls (282.82,50.58) and (282.65,51.54) .. (282.35,52.44) -- (272.81,49.58) -- cycle ; \draw   (272.81,59.07) .. controls (272.81,59.07) and (272.81,59.07) .. (272.81,59.07) .. controls (267.28,59.07) and (262.8,54.82) .. (262.8,49.58) .. controls (262.8,44.34) and (267.28,40.09) .. (272.81,40.09) .. controls (278.33,40.09) and (282.82,44.34) .. (282.82,49.58) .. controls (282.82,50.58) and (282.65,51.54) .. (282.35,52.44) ;  

\draw    (221.5,150.65) -- (256.09,168.37) ;
\draw [shift={(257.88,169.28)}, rotate = 207.11] [color={rgb, 255:red, 0; green, 0; blue, 0 }  ][line width=0.75]    (10.93,-3.29) .. controls (6.95,-1.4) and (3.31,-0.3) .. (0,0) .. controls (3.31,0.3) and (6.95,1.4) .. (10.93,3.29)   ;
\draw    (221.5,150.65) -- (256.13,131.25) ;
\draw [shift={(257.88,130.28)}, rotate = 150.74] [color={rgb, 255:red, 0; green, 0; blue, 0 }  ][line width=0.75]    (10.93,-3.29) .. controls (6.95,-1.4) and (3.31,-0.3) .. (0,0) .. controls (3.31,0.3) and (6.95,1.4) .. (10.93,3.29)   ;
\draw    (221.5,150.65) -- (255.88,150.3) ;
\draw [shift={(257.88,150.28)}, rotate = 179.41] [color={rgb, 255:red, 0; green, 0; blue, 0 }  ][line width=0.75]    (10.93,-3.29) .. controls (6.95,-1.4) and (3.31,-0.3) .. (0,0) .. controls (3.31,0.3) and (6.95,1.4) .. (10.93,3.29)   ;

\draw    (272.81,160.07) ;
\draw [shift={(275.62,160.91)}, rotate = 196.66] [fill={rgb, 255:red, 0; green, 0; blue, 0 }  ][line width=0.08]  [draw opacity=0] (8.93,-4.29) -- (0,0) -- (8.93,4.29) -- cycle    ;
\draw  [draw opacity=0] (272.81,160.07) .. controls (272.81,160.07) and (272.81,160.07) .. (272.81,160.07) .. controls (267.28,160.07) and (262.8,155.82) .. (262.8,150.58) .. controls (262.8,145.34) and (267.28,141.09) .. (272.81,141.09) .. controls (278.33,141.09) and (282.82,145.34) .. (282.82,150.58) .. controls (282.82,151.58) and (282.65,152.54) .. (282.35,153.44) -- (272.81,150.58) -- cycle ; \draw   (272.81,160.07) .. controls (272.81,160.07) and (272.81,160.07) .. (272.81,160.07) .. controls (267.28,160.07) and (262.8,155.82) .. (262.8,150.58) .. controls (262.8,145.34) and (267.28,141.09) .. (272.81,141.09) .. controls (278.33,141.09) and (282.82,145.34) .. (282.82,150.58) .. controls (282.82,151.58) and (282.65,152.54) .. (282.35,153.44) ;  

\draw    (38.5,72.65) -- (73.09,90.37) ;
\draw [shift={(74.88,91.28)}, rotate = 207.11] [color={rgb, 255:red, 0; green, 0; blue, 0 }  ][line width=0.75]    (10.93,-3.29) .. controls (6.95,-1.4) and (3.31,-0.3) .. (0,0) .. controls (3.31,0.3) and (6.95,1.4) .. (10.93,3.29)   ;
\draw    (38.5,72.65) -- (73.13,53.25) ;
\draw [shift={(74.88,52.28)}, rotate = 150.74] [color={rgb, 255:red, 0; green, 0; blue, 0 }  ][line width=0.75]    (10.93,-3.29) .. controls (6.95,-1.4) and (3.31,-0.3) .. (0,0) .. controls (3.31,0.3) and (6.95,1.4) .. (10.93,3.29)   ;
\draw    (38.5,72.65) -- (72.88,72.3) ;
\draw [shift={(74.88,72.28)}, rotate = 179.41] [color={rgb, 255:red, 0; green, 0; blue, 0 }  ][line width=0.75]    (10.93,-3.29) .. controls (6.95,-1.4) and (3.31,-0.3) .. (0,0) .. controls (3.31,0.3) and (6.95,1.4) .. (10.93,3.29)   ;

\draw    (89.81,82.07) ;
\draw [shift={(92.62,82.91)}, rotate = 196.66] [fill={rgb, 255:red, 0; green, 0; blue, 0 }  ][line width=0.08]  [draw opacity=0] (8.93,-4.29) -- (0,0) -- (8.93,4.29) -- cycle    ;
\draw  [draw opacity=0] (89.81,82.07) .. controls (89.81,82.07) and (89.81,82.07) .. (89.81,82.07) .. controls (84.28,82.07) and (79.8,77.82) .. (79.8,72.58) .. controls (79.8,67.34) and (84.28,63.09) .. (89.81,63.09) .. controls (95.33,63.09) and (99.82,67.34) .. (99.82,72.58) .. controls (99.82,73.58) and (99.65,74.54) .. (99.35,75.44) -- (89.81,72.58) -- cycle ; \draw   (89.81,82.07) .. controls (89.81,82.07) and (89.81,82.07) .. (89.81,82.07) .. controls (84.28,82.07) and (79.8,77.82) .. (79.8,72.58) .. controls (79.8,67.34) and (84.28,63.09) .. (89.81,63.09) .. controls (95.33,63.09) and (99.82,67.34) .. (99.82,72.58) .. controls (99.82,73.58) and (99.65,74.54) .. (99.35,75.44) ;

\draw (106.83,37.57) node [anchor=north west][inner sep=0.75pt]    {$ \begin{array}{l}
\mathbf{X}_{1} \ \ :=X_{1}^{( 0)}\\
\vdots \\
\mathbf{X}_{n_{1}^{M}} :=X_{n_{1}^{M}}^{( 0)}\\
\vdots \\
\mathbf{X}_{3n_{1}^{M}} :=X_{3n_{1}^{M}}{}_{\ }^{( 0)}
\end{array}$};
\draw (289.8,17.27) node [anchor=north west][inner sep=0.75pt]    {$ \begin{array}{l}
X_{1}^{( L)}\\
\vdots \\
X_{n_{2}^{L}}^{( L)}\\
\\
\\
\\
\vdots \\
X_{3n_{1}^{M} n_{2}^{L}}^{( L)}
\end{array}$};
\draw (268.1,26.5) node [anchor=north west][inner sep=0.75pt]  [font=\footnotesize]  {$L$};
\draw (268.1,127.5) node [anchor=north west][inner sep=0.75pt]  [font=\footnotesize]  {$L$};
\draw (1.5,61.4) node [anchor=north west][inner sep=0.75pt]    {$\mathbf{X}^{( 0)}$};
\draw (85.1,49.5) node [anchor=north west][inner sep=0.75pt]  [font=\footnotesize]  {$M$};

\end{tikzpicture}
        \caption{Amount of qubits in the composed protocol.}\label{fig:composed}
    \end{figure}
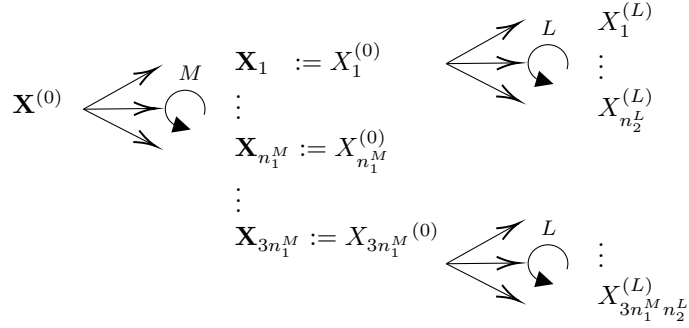
    Therefore, we achieve $\eps$-correctness whenever
    \[ \Pr(\left\{\textbf{X}^{(0)}\in S_{ECC}^\bot\setminus S_{ECC}\right\}\cup\left\{\sum_{j=n_1^M+1}^{3n_1^M}\omega(\textbf{X}_j)\geq 1\right\})\leq\eps. \]
    We can bound the probability of the first event in terms of the levels of concatenation of both the inner and outer code 
    \[ \Pr(\textbf{X}^{(0)}\in S_{ECC}^\bot\setminus S_{ECC})\leq \pth\left(\frac{\Pr(\textbf{X}_j^{(0)}\in S_{ECC}^\bot\setminus S_{ECC})}{\pth}\right)^{(t_1+1)^M}\leq\pth(p/\pth)^{(t_1+1)^L(t_1+1)^M}.  \]
    We can bound the probability of the second event with the union bound by requiring none of the traps to be triggered
    \[ \Pr(\sum_{j=n_1^M+1}^{3n_1^M}\omega(\textbf{X}_j)\geq 1)\leq 2n_1^M\pth(p/\pth)^{(t_1+1)^L}.  \]
    Note that if we fix the layers $L$ of the outer code, we cannot make the second equation as small as we desire by increasing the levels $M$ of concatenation of the inner code, i.e., correctness is determined by the size of the outer code as we expected.  We can rewrite $\eps$-correctness in terms of the qubits required for both the data and the traps to be protected
    \[ (t_2+1)^L(t_1+1)^M\geq\frac{\log(1/\eps)-\log(1/\pth)}{\log(\pth/p)}, \]
    and
    \[ \log(1/n_1^M)+(t_2+1)^L\log(\pth/p)\geq\log(1/\eps)-\log(1/2\pth).\]
    Substituting the inner layers $M$ sufficient to obtain $\delta$-security in~\cref{eq:sec_trap} in the above bounds we obtain
    \[ n_2^{L\log_{n_2}(t_2+1)}\gtrsim\log(1/\eps), \]
    and
    \[ n_2^{L\log_{n_2}(t_2+1)}\gtrsim\log(1/\eps)+\log(\log(1/\delta)^{\log_{2t_1+1}(n_1)}). \]
    
    In conclusion, the composed authentication and error-correcting code obtains $\eps$-correctness and $\delta$-security whenever the total amount of qubits grows as
    \[ 3 n_1^Mn_2^L\gtrsim \log(1/\eps)^{\log_{t_2+1}(n_2)}\log(1/\delta)^{\log_{2t_1+1}(n_1)}. \]
\end{proof}
\end{prop}

\subsection{Clifford scheme}

The Clifford code is a very efficient authentication scheme constructed as follows, see~\Cref{fig:clifford}. It form a set of purity testing codes by appending $n$ `traps' in the computational basis state $\dyad{0}$ to the data qubits; a Clifford operation to the resulting qubits in a random fashion attending to a secret shared key. The security of the Clifford group depends uniquely in the amount of traps that we append, this derives from the fact that the Clifford group not only maps Paulis to Paulis, but does so in a uniform distribution. Hence, the attacker has no control of the weight of the attack, as the Clifford twirl will map it into an arbitrary weight attack on the data and trap registers, and therefore with enough traps we can detect the attacks with very high probability.

\begin{figure}[h]
\begin{center}
    \begin{tabular}{ |c p{14cm}| }
    \hline
    \multicolumn{2}{|l|}{\textbf{Protocol 1}: Clifford authentication scheme, ``encode-then-encrypt'' form.} \\
    \hline
    \hline
    \multicolumn{2}{|l|}{\textbf{Encoding:}} \\
    1. & Alice and Bob obtain uniform keys $k$ (for the Cliffords) and $l$ (for the encryption) from the key resource.\\
    2. & Alice appends $n$ computational basis states $\dyad{0}$ to the message $\rho_A$. She applies a Clifford to all the qubit registers according to the secret key $k$.\\
    3. & Finally, she encrypts the message with a quantum one-time pad using the key $l$, obtaining thus
    \[ \sigma_{AS}=P_lC_k\left(\rho_A\otimes\dyad{0}^{\otimes n}\right)C_k^\dag P_l. \]\\
    4. & Alice sends $\sigma_{AS}$ to Bob through the insecure quantum channel.\\
    \hline
    \multicolumn{2}{|l|}{\textbf{Decoding:}} \\
    1. & Bob receives $\hat{\sigma}_{AS}$ and decrypts the data using $l$. Then he applies the inverse Clifford according to $k$ and measures the last $s$ registers in the computational basis respectively. If the measurement results in $\dyad{0}^n$, he accepts the protocol. Else, he aborts.\\
    \hline
    \end{tabular}
\end{center}
\caption{Clifford authentication scheme.}\label{fig:clifford}
\end{figure}

It is not difficult to show that the $n$-Clifford authentication scheme is $2^{-n}$-secure, but as for the trap scheme the protocol will reject whenever an error is present, making it impractical over noisy channels. However, we can compose the $n_1$-Clifford scheme with a $[[n_2^N,1,d_2^N]]$ error-correcting code, where once again each of the traps is also encoded in an error-correcting code. The same analysis as for the trap code yields that the composed protocol obtains $\eps$-correctness and $\delta$-security whenever the total amount of qubits grows as
\[ n_1n_2^N\gtrsim \log(1/\eps)^{\log_{t_2+1}(n_2)}\log(1/\delta). \]
\section{The threshold authentication scheme}

In this section we introduce the threshold scheme, an example of a quantum authentication scheme naturally robust against noisy channels. 

In Hayden, Leung and Mayers'~\cite{hayden_universal_2016} and Portmann's~\cite{portmann_quantum_2017} constructions of composed protocols, it is assumed that the authentication scheme rejects whenever an error is present -- which is always the case with very high probability when sending information through noisy channels -- and therefore an error-correcting code is necessary to make the schemes useful. However, from the structure of the composition, the number of qubits used in such a construction blows up both with the size of the purity-testing code used in the authentication scheme and the error-correcting code, as seen in the analysis of the trap scheme in~\Cref{prop:trap}. It is therefore natural to ask if such a composition is even necessary, and if we cannot design a protocol that directly constructs an authenticated quantum channel from a noisy insecure channel and shared secret key. This is exactly what the threshold scheme does.

The threshold scheme can be seen as an adaptation of the trap scheme where, with the same encoding, we require Bob to accept the message whenever a `low' amount of errors are detected. In other words, we use the traps as they were originally intended, to measure the amount of error present in the encoded data, and decide if these errors pertain to noise or an attack. The key idea is that we are double-encoding the data in two error-correcting codes, but this is not necessary as the outer error-correcting code in the trap scheme is to correct the noise, which we do directly with the inner code now. In principle this should not be enough as the correctable errors of an error-correcting code grow sublinearly with the size of the code, while for the depolarizing channel the number of errors is linear with the size. However, as we saw in~\Cref{sec:error}, concatenated codes correct linear amount of errors \emph{with very high probability}, which is enough to form a purity-testing family of codes. 

We will make these notions clear in the rest of the section. The threshold code is constructed as follows. Given a fixed $[[n,1,d]]$ error-correcting code, the scheme constructs a set of purity testing codes by appending $2n$ `traps' to the data qubits ($n$ computational-basis traps in the $\dyad{0}$ state and $n$ Hadamard-basis traps in the $\dyad{+}$ state); the resulting $3n$ qubit register is permuted in a random fashion attending to a secret shared key. In the decoding phase, after decryping and undoing the permutation, Bob accepts the protocol if less than a threshold $r=\alpha n$ errors are present in the traps. The threshold $r=\alpha n$ is a tuning of the amount of error that we are willing to accept in the traps without rejecting the authentication, assuming that these will be corrected by the error-correcting code. Hence, the parameter $\alpha$ depends on the noisy channel. The explicit construction of the threshold authentication scheme is given in~\Cref{fig:threshold}.

\begin{figure}[h]
\begin{center}
    \begin{tabular}{ |c p{14cm}| }
    \hline
    \multicolumn{2}{|l|}{\textbf{Protocol 1}: threshold authentication scheme $\pi_{AB}^{\text{thr}}$.} \\
    \hline
    \hline
    \multicolumn{2}{|l|}{\textbf{Encoding:}} \\
    1. & Alice and Bob agree on a $[[n,1,d]]$ quantum error-correcting code.\\
    2. & Alice and Bob obtain uniform keys $k$ (for the permutation) and $l$ (for the encryption) from the key resource.\\
    3. & Alice encodes the message $\rho_A$ with the agreed error-correcting code. Appends a $2n$ computational basis states $\dyad{0}^{\otimes 2n}$ and applies a Hadamard gate to the last $n$ qubits (so they are in the Hadamard-basis state $\dyad{+}$). She applies a permutation to all the qubit registers according to the secret key $k$.\\
    4. & Finally, she encrypts the message with a quantum one-time pad using the key $l$, obtaining thus
    \[ \sigma_{AS}=P_l\pi_k\left(\text{Enc}(\rho_A)\otimes\dyad{0}^{\otimes n}\otimes\dyad{+}^{\otimes n}\right)\pi^\dag_kP_l. \]\\
    5. & Alice sends $\sigma_{AS}$ to Bob through the noisy insecure quantum channel.\\
    \hline
    \multicolumn{2}{|l|}{\textbf{Decoding:}} \\
    1. & Bob receives $\hat{\sigma}_{AS}$ and decrypts the data using $l$. Then he applies the inverse permutation according to $k$ and measures the last $2n$ registers in the computational and Hadamard bases respectively. If less than a threshold $r=\alpha n$ of qubits differ from the expected outcome $\dyad{0}^n\otimes\dyad{+}^n$, he accepts the protocol. Else, he aborts.\\
    2. & If Bob accepts the protocol, he decodes the data register according to the agreed-upon error-correcting code.\\
    \hline
    \end{tabular}
\end{center}
\caption{Threshold authentication scheme.}\label{fig:threshold}
\end{figure}

Since there is no outer error-correcting code in our protocol, we have to ensure that the threshold scheme constructs a noiseless secure quantum channel from nothing but a noisy insecure quantum channel and a shared secret key. We will separate this task in two steps, first proving the correctness and then the security, according to~\Cref{def:security}.

\subsection{Correctness}

With correctness we mean that, in presence of no malicious player, Bob receives exactly the message that Alice sent as in~\Cref{fig:secure_noeve}. Therefore, we want to prove that when we use our protocol with a noisy channel, the outcome is nearly indistinguishable from using a noiseless secure channel without adversary. In other words, that the first condition of \Cref{def:security} holds.

\begin{figure}
\centering
    \begin{subfigure}{0.45\textwidth}
        \centering
        \begin{tikzpicture}[x=0.75pt,y=0.75pt,yscale=-1,xscale=1, scale=0.8]

\draw   (30.38,36.21) .. controls (30.38,32.89) and (33.07,30.2) .. (36.39,30.2) -- (54.43,30.2) .. controls (57.76,30.2) and (60.45,32.89) .. (60.45,36.21) -- (60.45,144.19) .. controls (60.45,147.51) and (57.76,150.2) .. (54.43,150.2) -- (36.39,150.2) .. controls (33.07,150.2) and (30.38,147.51) .. (30.38,144.19) -- cycle ;
\draw   (240.16,35.71) .. controls (240.16,32.39) and (242.85,29.7) .. (246.17,29.7) -- (264.22,29.7) .. controls (267.54,29.7) and (270.23,32.39) .. (270.23,35.71) -- (270.23,143.69) .. controls (270.23,147.01) and (267.54,149.7) .. (264.22,149.7) -- (246.17,149.7) .. controls (242.85,149.7) and (240.16,147.01) .. (240.16,143.69) -- cycle ;
\draw   (95.19,39.7) -- (205.81,39.7) -- (205.81,80.24) -- (95.19,80.24) -- cycle ;
\draw    (168.64,71.35) -- (234.98,71) ;
\draw [shift={(237.98,70.98)}, rotate = 179.7] [fill={rgb, 255:red, 0; green, 0; blue, 0 }  ][line width=0.08]  [draw opacity=0] (8.93,-4.29) -- (0,0) -- (8.93,4.29) -- cycle    ;
\draw   (130.3,54.25) -- (168.51,54.25) -- (168.51,75.14) -- (130.3,75.14) -- cycle ;

\draw    (130.44,71.1) -- (65.76,70.92) ;
\draw [shift={(62.76,70.91)}, rotate = 0.16] [fill={rgb, 255:red, 0; green, 0; blue, 0 }  ][line width=0.08]  [draw opacity=0] (8.93,-4.29) -- (0,0) -- (8.93,4.29) -- cycle    ;

\draw    (240.34,49.96) -- (193.01,49.99) ;
\draw [shift={(191.01,49.99)}, rotate = 359.96] [color={rgb, 255:red, 0; green, 0; blue, 0 }  ][line width=0.75]    (10.93,-3.29) .. controls (6.95,-1.4) and (3.31,-0.3) .. (0,0) .. controls (3.31,0.3) and (6.95,1.4) .. (10.93,3.29)   ;
\draw    (60.67,49.96) -- (108.02,49.99) ;
\draw [shift={(110.02,49.99)}, rotate = 180.04] [color={rgb, 255:red, 0; green, 0; blue, 0 }  ][line width=0.75]    (10.93,-3.29) .. controls (6.95,-1.4) and (3.31,-0.3) .. (0,0) .. controls (3.31,0.3) and (6.95,1.4) .. (10.93,3.29)   ;
\draw   (99.96,108.77) -- (200.76,108.77) -- (200.76,150.7) -- (99.96,150.7) -- cycle ;
\draw    (54.37,129.63) -- (130.74,129.43) -- (130.8,180.35) ;
\draw    (20.67,129.46) -- (36.93,129.53) ;
\draw    (265.77,129.64) -- (278.37,129.7) -- (279.03,129.7) ;
\draw [shift={(282.03,129.71)}, rotate = 180.25] [fill={rgb, 255:red, 0; green, 0; blue, 0 }  ][line width=0.08]  [draw opacity=0] (8.93,-4.29) -- (0,0) -- (8.93,4.29) -- cycle    ;
\draw    (244.58,129.75) -- (169.56,129.68) -- (169.82,180.05) ;
\draw   (115.11,173.73) .. controls (115.11,171.39) and (117,169.5) .. (119.34,169.5) -- (180.88,169.5) .. controls (183.22,169.5) and (185.11,171.39) .. (185.11,173.73) -- (185.11,186.42) .. controls (185.11,188.75) and (183.22,190.65) .. (180.88,190.65) -- (119.34,190.65) .. controls (117,190.65) and (115.11,188.75) .. (115.11,186.42) -- cycle ;
\draw    (160.94,180.03) .. controls (161.04,174.73) and (164.74,174.33) .. (165.64,180.03) .. controls (166.54,185.73) and (169.74,185.85) .. (169.82,180.05) ;
\draw    (130.8,180.35) .. controls (130.64,187.63) and (134.79,185.22) .. (135.04,180.03) .. controls (135.3,174.85) and (139.42,174.02) .. (139.34,180.33) ;
\draw    (152.54,180.23) .. controls (152.64,174.93) and (156.04,172.73) .. (156.84,180.03) .. controls (157.64,187.33) and (160.87,185.84) .. (160.94,180.03) ;
\draw    (139.34,180.33) .. controls (139.19,187.62) and (143.09,185.42) .. (143.34,180.23) .. controls (143.6,175.05) and (148.02,173.82) .. (147.94,180.13) ;
\draw    (147.94,180.13) .. controls (147.94,182.83) and (149.24,185.33) .. (150.34,185.33) .. controls (151.44,185.33) and (152.54,183.03) .. (152.54,180.23) ;

\draw (239.38,10.99) node [anchor=north west][inner sep=0.75pt]   [align=left] {Bob};
\draw (25.7,12.07) node [anchor=north west][inner sep=0.75pt]   [align=left] {Alice};
\draw (133.9,56.65) node [anchor=north west][inner sep=0.75pt]   [align=left] {Key};
\draw (67.06,31.4) node [anchor=north west][inner sep=0.75pt]   [align=left] {{\small req.}};
\draw (210.56,31.5) node [anchor=north west][inner sep=0.75pt]   [align=left] {{\small req.}};
\draw (34.75,34.9) node [anchor=north west][inner sep=0.75pt]    {$\pi _{A}$};
\draw (244.55,34.4) node [anchor=north west][inner sep=0.75pt]    {$\pi _{B}$};
\draw (144.27,111.4) node [anchor=north west][inner sep=0.75pt]    {$\mathcal{C}$};
\draw (284.05,122.7) node [anchor=north west][inner sep=0.75pt]    {$\rho ,\ \bot $};
\draw (4,123.33) node [anchor=north west][inner sep=0.75pt]    {$\rho $};
\draw (77.85,73.4) node [anchor=north west][inner sep=0.75pt]    {$k$};
\draw (211.85,73.8) node [anchor=north west][inner sep=0.75pt]    {$k$};
\draw (189.2,182) node [anchor=north west][inner sep=0.75pt]    {$\#$};

\end{tikzpicture}
        \caption{Threshold protocol with no adversary present.}
    \end{subfigure}\hfill
    \begin{subfigure}{0.45\textwidth}
        \centering
        \begin{tikzpicture}[x=0.75pt,y=0.75pt,yscale=-1,xscale=1,scale=0.9]

\draw   (60.38,19.99) -- (220.88,19.99) -- (220.88,81.49) -- (60.38,81.49) -- cycle ;
\draw    (168.17,56.26) -- (168.09,110.74) ;
\draw  [dash pattern={on 4.5pt off 4.5pt}]  (111.28,50.61) -- (111.37,107.99) ;
\draw [shift={(111.38,109.99)}, rotate = 269.91] [color={rgb, 255:red, 0; green, 0; blue, 0 }  ][line width=0.75]    (10.93,-3.29) .. controls (6.95,-1.4) and (3.31,-0.3) .. (0,0) .. controls (3.31,0.3) and (6.95,1.4) .. (10.93,3.29)   ;
\draw    (40.51,50.12) -- (149.88,49.99) -- (180.38,59.99) ;
\draw    (180.34,50.55) -- (246.68,50.2) ;
\draw [shift={(249.68,50.18)}, rotate = 179.7] [fill={rgb, 255:red, 0; green, 0; blue, 0 }  ][line width=0.08]  [draw opacity=0] (8.93,-4.29) -- (0,0) -- (8.93,4.29) -- cycle    ;
\draw    (171.26,56.76) -- (171.21,107.41) -- (171.21,110.77) ;
\draw   (94.5,114.3) .. controls (94.5,111.92) and (96.42,110) .. (98.8,110) -- (181.48,110) .. controls (183.86,110) and (185.78,111.92) .. (185.78,114.3) -- (185.78,127.2) .. controls (185.78,129.57) and (183.86,131.5) .. (181.48,131.5) -- (98.8,131.5) .. controls (96.42,131.5) and (94.5,129.57) .. (94.5,127.2) -- cycle ;

\draw (90.9,90.05) node [anchor=north west][inner sep=0.75pt]    {$m$};
\draw (175.66,89.61) node [anchor=north west][inner sep=0.75pt]    {$0$};
\draw (222.88,22.99) node [anchor=north west][inner sep=0.75pt]   [align=left] {Bob};
\draw (255.51,43.2) node [anchor=north west][inner sep=0.75pt]    {$\rho ,\ \bot $};
\draw (22.2,22.57) node [anchor=north west][inner sep=0.75pt]   [align=left] {Alice};
\draw (23.5,46) node [anchor=north west][inner sep=0.75pt]    {$\rho $};
\draw (189.5,122.4) node [anchor=north west][inner sep=0.75pt]    {$\diamondsuit $};

\end{tikzpicture}
        \caption{Authenticated quantum channel with no adversary present.}\label{fig:secure_noeve}
    \end{subfigure}
\caption{Comparison between the threshold protocol and a secure authenticated quantum channel without adversary.}
\end{figure}
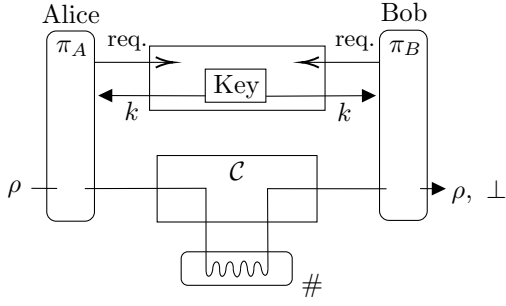
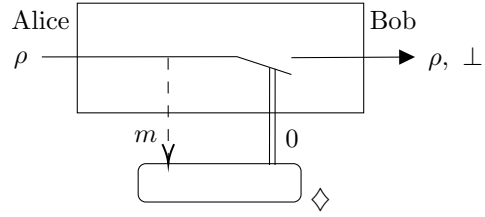

\begin{prop}\label{prop:thr_correct}
Let $\#_E$ be a filter introducing the noise given by the depolarizing channel with channel error $p<3\pth/4$. The threshold authentication scheme $\pi_{AB}^{\text{trh}}$ with an $[[n_1^M,1,d_1^M]]$ inner code and threshold parameter $\alpha>4p/3$ is $\eps$-correct, i.e.
\[ d(\pi_{AB}^{\text{thr}}(\mathcal{C}_\#||\mathcal{K}),\mathcal{S}_\Diamond)\leq\eps, \]
with
\[ \eps=\pth(p/\pth)^{(t_1+1)^M}+\exp(-n(\alpha-4p/3)^2). \]
    \begin{proof}
    For simplicity we will denote $n=n_1^M$. To prove correctness within $\eps$ we have to show that the threshold protocol $\pi^{\text{thr}}_{AB}$ constructs a noiseless secure channel $\mathcal{S}_\Diamond$ such that the real system transmitted through a noisy channel $\pi_{AB}^{\text{thr}}(\mathcal{C}_\#||\mathcal{K})$ cannot be distinguished from the ideal system $\mathcal{S}_\Diamond$. Note that distinguishability in presence of no adversary is exactly the diamond norm between the identity map and the encoding-noise-decoding map of the threshold code, i.e.
    \[ d(\pi_{AB}^{\text{thr}}(\mathcal{C}_\#||\mathcal{K}),\mathcal{S}_\Diamond)=\frac{1}{2}\left\|\mathcal{D}^{\text{thr}}\circ\mathcal{F}\circ\mathcal{E}^{\text{thr}}-I\right\|_\diamond. \]
    Let us denote by $X_1,\ldots, X_{3n}$ the independent random variables such that the first $n$ fail with probability $p$, i.e., $\Pr(\omega(X_j)=1)=p$ for all $j=1,\ldots,n$, and the last $2n$ fail with probability $2p/3$, i.e., $\Pr(\omega(X_j)=1)=2p/3$ for all $j=n,\ldots 3n$. Let us denote by $\textbf{X}$ the tensor product of the first $n$ variables, $\textbf{X}:=X_1\otimes\cdots\otimes X_n$. We do this distinction because we have computational basis and Hadamard basis traps, thus the probability of rejection is different. We achieve $\eps$-correctness whenever the rejection probability of the traps or the failed recovery of the error-correcting code encoding the data qubits is less than $\eps$. That is,
    \[ \Pr\left(\left\{\textbf{X}\in S^\bot\setminus S\right\}\cup\left\{\sum_{j=n}^{3n}\omega(X_j)\geq\alpha n\right\}\right)=\Pr\left(\textbf{X}\in S^\bot\setminus S\right)+\Pr\left(\sum_{j=n}^{3n} \omega(X_j)\geq\alpha n\right)\leq\eps. \]
    
    On the one hand, by the analysis of concatenated codes in~\Cref{sec:error}, whenever the error of the depolarizing channel is smaller than the threshold value
    \[ p< p_{\text{thr}}:=\binom{n}{t+1}^{-1}, \]
    we can make the rejection probability of the encoded data as small as desired by increasing the levels of concatenation
    \[ \Pr\left(\textbf{X}\in S^\bot\setminus S\right)\leq p_{\text{thr}}\left(p/p_{\text{thr}}\right)^{(t_1+1)^M}. \]
    On the other hand, although the traps undergo an independent identically distributed (i.i.d.) noise model, the decoding will only reject whenever a threshold $r=\alpha n$ of them are triggered. By Hoeffding's inequality
    \begin{equation}\begin{split} \Pr\left(\sum_{j=n}^{3n} \omega(X_j)\geq r\right)&=\Pr\left(\sum_{j=n}^{3n}\omega(X_j)-2n\frac{2p}{3}\geq \alpha n-2n\frac{2p}{3}\right)\leq\exp(-2\frac{\left(\left(\alpha-\frac{4p}{3}\right)n\right)^2}{2n})\\
    &=\exp(-n\left(\alpha-4p/3\right)^2), \end{split}\end{equation}
    whenever $\alpha>4p/3$.
    \end{proof}
\end{prop}

\subsection{Security}

Recall that with security we mean that in presence of a malicious player, there exists a simulator in the `ideal protocol' that is indistinguishable from the `real protocol'. In other words, that the second condition of \Cref{def:security} is satisfied. However, instead of constructing this simulator, it is enough to show that the threshold scheme constructs a set of codes that is purity testing, which will provide us with security by~\Cref{thm:sec_purity}. Although Portmann's original proof constructs a secure channel from a noiseless channel, in the security proof the filters are substituted by an adversary and therefore work for our setting as well. 

The idea of the security analysis is to note that in the previous security analysis the inner error correcting codes were used to detect errors of low weight, which are sublinear in the size of the protocol. But more is true: if we look at the concatenated codes from~\Cref{sec:error}, we see that they actually correct a linear amount of errors with very high probability. By setting the threshold properly, we can exploit this fact to prove that the threshold scheme is purity testing.

\begin{prop}
    The set of purity testing codes described by the threshold code with an inner $[[n_1^M,1,d_1^M]]$ concatenated error-correcting code and threshold parameter $\alpha<p_{\text{thr}}$, indexed by the key for the permutation $k\in\mathcal{K}$, is $\delta$-purity testing, where 
    \begin{equation}\label{eq:thr_security} \delta=\max\left\{\pth \frac{9(\alpha/\pth)^{(t_1+1)^M}}{10\sqrt{6\pi n\alpha (1-\alpha)}},\exp(-\frac{\alpha n}{4})\right\}. \end{equation}
    \begin{proof}
    For simplicity we will denote $n=n_1^M$. The threshold code, for a key $k\in\mathcal{K}$, is characterized by unitaries $V_k=\pi_k(\mathrm{Enc}\otimes I^{\otimes n}\otimes H^{\otimes n})$ and syndrome $\dyad{0}^{\otimes(n-1)}\otimes\dyad{s}^{\otimes 2n}$, where
    \[ \dyad{s}^{\otimes 2n}=\frac{1}{(2n)!2^r}\sum_{\pi\in\Pi_{2n}}\pi^\dag(I^{\otimes r}\otimes\dyad{0}^{\otimes(2n-r)})\pi. \]
    The first $n-1$ syndromes are used to decode the inner error-correcting code, and the last $2n$ for the traps such that the protocol rejects whenever more than $r$ non-zero traps are detected. Moreover, let us denote by $\{S_k\}$ the keyed stabilizer subgroups and by $S_{\text{ECC}}$ the stabilizer subgroup of the inner error-correcting code. For a particular permutation $\pi_k$, on the one hand, the set of Paulis that are not detected is
    \[ S_k^\bot =\{ \pi_k^\dag(P\otimes Q\otimes R)\pi_k\colon P\in S_{\text{ECC}}^\bot,\:\omega_X(Q)+\omega_Z(R)\leq r\}. \]
    On the other hand, since the traps are invariant to $Z$ and $X$ operations respectively, the Paulis that act trivially on the message are
    \[ S_k =\{ \pi_k^\dag(P\otimes Q\otimes R)\pi_k\colon P\in S_{\text{ECC}},\: Q\in\{I,Z\}^{\otimes n},\: R\in\{I,X\}^{\otimes n}\}. \]
    
    To prove that the threshold code is $\delta$-purity testing we have to show that if the permutation key is selected uniformly at random, the probability of any Pauli error $E\in\mathcal{G}_n$ acting non-trivially on the data and not being detected is upper bounded by $\delta$. Although the attacker can fix the weight of the attack, the chosen Pauli operation is irrelevant, as the secret permutation makes it looks like a $X$ or $Z$ Pauli with equal probability. We will now divide the proof in two cases attending to the weight $\omega:=\omega(E)$. First we split the set of permutations in terms of the error correction and the trap detection
    \begin{equation}\begin{split}
    \Pi^0(E) &:= \{\pi\in\Pi_{3n}\colon E=\pi^\dag(P\otimes T)\pi,\: P\in S_{\text{ECC}}^\bot\setminus S_{\text{ECC}},\: T\in\mathcal{G}_{2n}\},\\
    \Pi^1(E) &:= \{\pi\in\Pi_{3n}\colon E=\pi^\dag(P\otimes Q\otimes R)\pi,\: P\in \mathcal{G}_n,\: \omega_X(Q)+\omega_Z(R)\leq r\},
    \end{split}\end{equation}
    so that we can bound the purity testing parameter by the minimum size of both sets
    \begin{equation}\label{eq:purity_thr}\Pr_{k\in\mathcal{K}}\left(E\in S_k^\bot\setminus S_k\right)\leq\min\left\{\frac{\left|\Pi^0(E)\right|}{\left|\Pi_{3n}\right|},\frac{\left|\Pi^1(E)\right|}{\left|\Pi_{3n}\right|}\right\}. \end{equation}
    Consequently, it is enough to bound one of the sets for different weight attacks.
    
    \underline{Case 1}: $\omega\leq 3r$. For low-weight attacks, still linear in the total size of the protocol, we expect the error-correcting code to correct them with high probability, see the discussion in~\Cref{sec:error}. Since the set of Pauli operators acting non-trivially and being undetected is exactly the one that the error-correcting code fails to decode correctly, we can rewrite it in terms of random variables. Let us define the following set of i.i.d.\ random variables $X_1,\ldots, X_{3n}$ such that $X_j\in\{X,Y,Z\}$ with probability $\omega/3n$ and $X_j=I$ with probability $1-\omega/3n$. Let us denote by $\textbf{X}$ the tensor product of the first $n$ variables, $\textbf{X}:=X_1\otimes\cdots\otimes X_n$. Then the weight of these variables is
    \begin{equation}
        \omega(X_j):=
        \begin{cases} 1 & \text{if $X_j\not=I$}\\
        0 & \text{otherwise}  \end{cases},
        \quad\text{with}\quad \Pr(\omega(X_j)=1)=\frac{\omega}{3n}.
    \end{equation}
    Now if we condition on a fixed amount of registers suffering an error, we have the bound
    \[ \frac{\left|\Pi^0(E)\right|}{\left|\Pi_{3n}\right|}\leq \Pr(\textbf{X}\in S_{\text{ECC}}^\bot\setminus S_{\text{ECC}}\left|\sum_{j=1}^{3n} \omega(X_j)=\omega\right.)\leq\frac{\Pr\left(\textbf{X}\in S_{\text{ECC}}^\bot\setminus S_{\text{ECC}}\right)}{\Pr(\sum_{j=1}^{3n} \omega(X_j)=\omega)}. \]
    The numerator in the LHS is exactly the probability of failed recovery of an error-correcting code when the message is sent through a channel with error probability $\omega/{3n}$ per qubit, which is bounded by~\cref{eq:threshold}. The denominator is just the probability of the binomial distribution having the expected value and is lower bounded by 
    \[ \Pr(\sum_{j=1}^{3n} \omega(X_j)=\omega)\geq\frac{9}{10}\frac{1}{\sqrt{2\pi\omega(1-\omega/3n)}}, \]
    see~\Cref{app:binom}. Therefore, 
    \begin{equation} \frac{\left|\Pi^0(E)\right|}{\left|\Pi_{3n}\right|}\leq \pth \frac{9(\omega/3n\pth)^{(t_1+1)^M}}{10\sqrt{2\pi\omega(1-\omega/3n)}}, \end{equation}
    where $\pth=\binom{n_1}{t_1+1}^{-1}$ is the threshold of the error-correcting code. We can make the above bound as small as desired by increasing the levels $M$ of concatenation whenever $\omega<3n\pth$, which holds since $\alpha<\pth$ by hypothesis.

    \underline{Case 2}: $\omega\geq3r$. High weight attacks will be detected by the traps with high probability, even when a linear amount of them $r=\alpha n$ are triggered before aborting the protocol. Although we cannot exploit the idependence of errors as in the security proof of the trap code, we can apply a sampling variant of the Chernoff bound, see~\Cref{app:chernoff}. Let us define the total population to be all the registers $A:=\{1,\ldots,3n\}$, and the sub-population the traps $B:=\{n,\ldots,3n\}$. Given a Pauli attack $E$ of weight $\omega=\omega(E)$ and a random sample $S\subset A$, with $|S|=\omega$, we can define
    \begin{equation}
        \omega(X_j):=
        \begin{cases} 1 & \text{if the $j$-th register suffers an error, } E_j\in\{X,Y,Z\},\\
        0 & \text{if } E_j=I.  \end{cases}
    \end{equation}
    In this case, the probability over the key of the attack not being detected by the traps is equivalent to the probability of the relative size of the traps in the sampling being below the threshold, since otherwise they will be detected. We can write this explicitly
    \begin{equation}\begin{split} \Pr(\sum_{j=n}^{3n}\omega(X_j)<r)&=\Pr(\sum_{j=n}^{3n}\omega(X_j)<(1-\gamma)\frac{|B|}{|A|}\omega)<\exp(-\gamma^2\frac{|B|}{|A|}\frac{\omega}{2})\\
    &= \exp(-\frac{\omega}{3}\left(1-\frac{3r}{2\omega}\right)^2),
    \end{split}\end{equation}
    where $\gamma=1-\frac{3r}{2\omega}$, with $\gamma\in(0,1)$ whenever $\omega>3r/2$.

    Of course the attacker will choose the best possible attack for a given protocol, thus we need a bound independent of the weight. Since the probability of rejection of the error-correcting code is increasing in the size of the protocol, while the probability of the traps being triggered is decreasing, we can bound~\cref{eq:purity_thr} by considering the worst attack for the error correction $\omega=3r$ in the detection as well.
    \[ \Pr_{k\in\mathcal{K}}\left(E\in S_k^\bot\setminus S_k\right)\leq\max\left\{\pth \frac{9(\alpha/\pth)^{(t_1+1)^M}}{10\sqrt{6\pi n\alpha (1-\alpha)}},\exp(-\frac{\alpha n}{4})\right\}. \]
    \end{proof}
\end{prop}

The above proposition in combination with~\Cref{thm:sec_purity} is enough to prove security of the threshold code. This is clear from the splitting of correctness and security parameters in~\Cref{def:security} because \emph{security} refers to the comparison of real and ideal protocols in presence of an adversary, and therefore the channels are compared without filters. The following theorem is therefore a direct consequence.

\begin{theorem}\label{thm:thr_sec}
Let $\#_E$ be a filter introducing the noise given by the depolarizing channel with channel error $p<3\pth/4$. The threshold authentication scheme $\pi_{AB}^{\text{thr}}$ with an $[[n^M,1,d^M]]$ inner code and threshold parameter $\alpha<p_{\text{thr}}$ is $\delta$-secure, i.e. there exists a converter $\sigma_E$ such that
\[ d(\pi_{AB}^{\text{thr}}(\mathcal{C}||\mathcal{K}),\sigma_E\mathcal{S})\leq\delta, \]
with
\[ \delta = \max\left\{\pth \frac{9(\alpha/\pth)^{(t_1+1)^M}}{10\sqrt{6\pi n\alpha (1-\alpha)}},\exp(-\frac{\alpha n}{4})\right\}. \]
\end{theorem}

\subsection{Efficiency in terms of qubits}

We can combine the correctness and security requirements of the threshold scheme to obtain the sufficient amount of qubits that the threshold scheme requires to obtain $(\eps,\delta)$-security. Although we studied both parameters separately, the effectiveness of the threshold scheme lies in the fact that the size of the inner error-correcting code determines both the correctness and security. That is, in contrast to the composed authentication and error correction, we can construct a secure quantum channel from a noisy insecure channel and secret key without the need to double encode our qubits in two error-correcting codes.

\begin{theorem}
Let $\#_E$ be a filter introducing the noise given by the depolarizing channel with channel error $p<3\pth/4$. The threshold authentication scheme $\pi_{AB}^{\text{thr}}$ with an $[[n_1^M,1,d_1^M]]$ inner code and threshold parameter $\alpha\in\left(\frac{4p}{3},\pth\right)$ to obtain $\eps$-correctness and $\delta$-security, i.e.
\[ \hat{\mathcal{C}}_\#||\mathcal{K}\xrightarrow{\pi^{\text{thr}},(\eps,\delta)}\mathcal{S}_\Diamond, \]
it is sufficient for the amount of qubits to grow as
\[ O\left(\max\left\{\log(1/\eps)^{C},\log(1/\delta)^{C}\right\}\right). \]
Here the constant $C(t_1,n_1):=\log_{t_1+1}(n_1)$ depends only on the properties of the error-correcting code chosen for the concatenation.
    \begin{proof}
    We will start by writing the correctness in terms of the number of qubits required. By~\Cref{prop:thr_correct}, the threshold code will obtain $\eps$-correctness whenever
    \[ \pth(p/\pth)^{(t_1+1)^M}\leq \eps/2,\quad\text{and}\quad \exp(-n(\alpha-4p/3)^2)\leq\eps/2. \]
    Since the scaling of the error correction is faster than the scaling of the trap detection, bounding the former will be enough. More precisely, 
    \[ n^{\log_{n_1}(t_1+1)}\geq\frac{\log(2/\eps)-\log(1/\pth)}{\log(\pth/p)}.\]
    To study the security we can bound the values from~\Cref{thm:thr_sec}, i.e.
    \[ \pth \frac{9(\alpha/\pth)^{(t_1+1)^M}}{10\sqrt{6\pi n\alpha (1-\alpha)}}\leq \delta,\quad\text{and}\quad \exp(-\frac{\alpha n}{4})\leq\delta. \]
    Once again, since the scaling of the qubits contributed by the error correction is faster, it is enough to bound
    \[ n^{\log_{n_1}(t_1+1)}\log(\pth/\alpha)+\frac{1}{2}\log(n)\geq\log(1/\delta)-\log(\frac{10\sqrt{6\pi\alpha(1-\alpha)}}{9\pth}).\]
    Ignoring the precise multiplying constants, we obtain that the order of growth sufficient for the threshold code to obtain $\eps$-correctness and $\delta$-security is
    \[ 3n_1^M\gtrsim\max\left\{\log(1/\eps)^{\log_{t_1+1}(n_1)},\log(1/\delta)^{\log_{t_1+1}(n_1)}\right\}. \]
    \end{proof}
\end{theorem}

\section{Conclusions}

We studied the combination of authentication and error-correction in a single primitive, and we saw that  the size blowups of authentication and error-correction are (slightly-more than) \emph{multiplied} in the composed protocol to determine the total blowup. As an example of the potential of looking at these properties together, we designed the threshold scheme, for which the resource usage is only dependent on the maximum blowup of the two functionalities. In particular, when comparing the trap scheme with the threshold scheme for any i.i.d.~noise channel, the best error-correcting code for the outer error correction in the composed protocol can always be employed as the inner code for the threshold scheme, such that we get a guaranteed polynomial-order improvement in the amount of qubits that suffice to achieve the same level of security and robustness.

We leave as an open question what is the maximum gain in efficiency of combining these functionalities. For instance, the Clifford code is a more efficient authentication scheme than the trap code, however the threshold code performs better in many ranges of parameters than the Clifford code with error-correction wrapped around it. In particular, for equal values of correctness and security, as considered in the previous cryptographic security definitions~\cite{portmann_quantum_2017, maurer_abstract_2011}, we always obtain a polynomial-order improvement from the composed protocol. This invites the interesting question of whether it is possible to make the Clifford code error-robust in a more efficient way, or whether the threshold code is the most-efficient combination possible. Our analysis opens the door to combining more general error-correction and authentication codes, which could improve the practicality of the resulting scheme. 

To summarize, we include a comparison between the different authentication schemes mentioned in the article with some existing error-correcting code parameters in~\Cref{table:summary}. Here we consider the $\eps$-cryptographic security, i.e.\ $\eps$-correctness and $\eps$-security, and note how we always get a polynomial improvement in the amount of qubits required.
\begin{figure}[H]
    \begin{center}
    \begin{tabular}{|c|c|c|c|}
         \hline
         & $[[n^L,1,d^L]]$-code & $[[7^L,1,3^L]]$-Steane & $[[5^L,1,3^L]]$-code\\
         \hline
         Trap code with inner and outer & $\log_{t+1}(n)+\log_{2t+1}(n)$ & $4.58$ & $3.79$\\
         \hline
         Clifford code with outer & $1+\log_{t+1}(n)$ & $3.81$ & $3.32$\\
         \hline
         Threshold code with inner & $\log_{t+1}(n)$ & $2.81$ & $2.32$\\
         \hline
    \end{tabular}
    \end{center}
    \caption{Growth exponents of $\log(1/\eps)$ for different authentication codes.}\label{table:summary}
\end{figure}

In the current work we only considered the notions of information-theoretic security where the integrity of the plaintext is important, and we do not study key recycling.
It could be interesting to combine some of these notions -- for instance, to construct a computationally-secure scheme for authentication which also functions as error-correcting code in an efficient way.

Additionally, a code which is both error-correcting and authenticating is in some sense the \emph{opposite} of ciphertext authentication. Therefore it could be interesting to consider if there is a natural way of combining these functionalities, and what the maximum amount of key recycling possible is.

\addcontentsline{toc}{section}{References}
\printbibliography

\appendix
\section{Useful theorems}

The following variant of Chernoff's bound studies the probability of the majority in a population becoming the minority, and vice versa.
\begin{lemma}[{\cite[Lemma 1]{goldberg_competitive_2001}}]\label{app:chernoff} Consider a population set $A$ and a sub-population $B\subset A$. Suppose we pick an integer $k$ such that $0<k<|A|$ and a random subset $S\subset A$ of size~$k$. Then for any $0<\gamma\leq 1$ we can bound the relative size of the sub-population in the sample $S$ by
	\[ P\left[\frac{|S\cap B|}{k}<(1-\gamma)\frac{|B|}{|A|}\right]<\exp\left(-\gamma^2\frac{|B|}{|A|}\frac{k}{2}\right). \]
\end{lemma}

Although the probabilities of the binomial distribution are well known, it is sometimes easier to give a tractable lower bound. Here we derive one for the probability of a binomially distributed random variable attaining its mean by applying the Stirling's approximation.
\begin{lemma}\label{app:binom} 
Let $X\sim B(n,p)$, with $n\geq3$ and $np\geq 1$. Then the probability of $X$ obtaining its expected value, $\mathbb{E}(X)=np$, is lower bounded by
\[ \Pr(X=np)\geq \frac{9}{10}\frac{1}{\sqrt{2\pi np(1-p)}}. \]
\begin{proof}
The bound follows from Robbins~\cite{robbins_remark_1955} version of Stirling's approximation. This is, for every $n\geq 1$, it holds that
\begin{equation}\label{eq:robbin}
\sqrt{2\pi n}\left(\frac{n}{e}\right)^n e^{\frac{1}{12n+1}}<n!<\sqrt{2\pi n}\left(\frac{n}{e}\right)^n e^{\frac{1}{12n}}. \end{equation}
The binomial coefficient is nothing but a fraction of factorials 
\[ \binom{n}{np}=\frac{n!}{(np)!(n-np!)}, \]
thus substituting the LHS of~\cref{eq:robbin} for the numerator and the RHS for the denominator in the above equation, we obtain
\begin{equation}\begin{split}
\binom{n}{np}p^{np}(1-p)^{n-np} & \geq p^{np}(1-p)^{n-np} \frac{\sqrt{2\pi n}}{\sqrt{2\pi np}\sqrt{2\pi(n-np)}} \frac{n^n e^{np}e^{n-np}}{e^n(np)^{np}(n-np)^{n-np}} \frac{e^{\frac{1}{12n+1}}}{e^{\frac{1}{12np}}e^{\frac{1}{12(n-np)}}}\\
& = \frac{1}{\sqrt{2\pi np(1-p)}}f(n,p),
\end{split}\end{equation}
where 
\[ f(n,p):=\frac{e^{\frac{1}{12n+1}}}{e^{\frac{1}{12np}}e^{\frac{1}{12(n-np)}}}. \]
Since this function is a symmetric on $p$ and increasing on $n$, we can lower bound it by its extreme values
\[ f(n,p)\geq f(3,1/3)\geq \frac{9}{10}. \]
\end{proof}
\end{lemma}

\section{Security of the trap code}\label{app:trap}

In order to be faithful to the trap code, we apply the same thorough security analysis in terms of the inner error-correcting code. Note that we are trying to exploit the error-correcting capabilities of the inner error-correcting code, which corrects a linear amount of errors with high probability. However, turns out that the detection probability of the traps scales faster than the correction of the error-correction, and therefore we cannot push the intersection further than the linear distance mark.

\begin{prop}
The trap code with inner concatenated error-correcting code $[[n_1^M,1,d_1^M]]$ is $\delta$-purity testing, where
\[\delta=\max_{\substack{\omega\geq t}}\left\{\min\left\{(2/3)^{\omega/2},\pth \frac{9(\omega/3n\pth)^{(t_1+1)^M}}{10\sqrt{2\pi\omega(1-\omega/3n)}}\right\}\right\}.\]
\begin{proof}
For simplicity we will denote $n=n_1^M$. The trap code, for a key $k\in\mathcal{K}$, is characterized by unitaries $V_k= \pi_k(\text{Enc}\otimes I^{\otimes n}\otimes H^{\otimes n})$ and syndrome $\dyad{0}^{\otimes (3n-1)}$. The first $n-1$ syndromes are used to decode the inner error-correcting code, and the last $2n$ for the traps. Moreover, let us denote by $\{S_k\}$ the keyed stabilizer subgroups and by $S_{\text{ECC}}$ the stabilizer subgroup of the inner error-correcting code. For a particular permutation $\pi_k$, the set of Paulis that are not detected is
\[ S_k^\bot = \{ \pi_k^\dag (P\otimes Q\otimes R)\pi_k\colon P\in S_{\text{ECC}}^\bot,\: Q\in\{I,Z\}^{\otimes n},\: R\in\{I,X\}^{\otimes n} \}, \]
i.e. the ones that are not detected neither by the inner error-correcting code nor the traps. Similarly, since the traps are invariant with respect to $Z$ and $X$ operations respectively, the Paulis that act trivially on the message are 
\[ S_k = \{ \pi_k^\dag (P\otimes Q\otimes R)\pi_k\colon P\in S_{\text{ECC}},\: Q\in\{I,Z\}^{\otimes n},\: R\in\{I,X\}^{\otimes n} \}. \]
In order to show that the trap code is purity testing, let $E\in\mathcal{G}_{3n}$ be a Pauli error of fixed weight $\omega(E)=\omega$, the probability of it acting non-trivially on the data and not being detected is
\begin{equation}\label{eq:prob_long_perm} \Pr_{k\in\mathcal{K}}\left(E\in S_k^\bot\setminus S_k\right)=\frac{\left| \{\pi\colon E=\pi^\dag(P\otimes Q\otimes R)\pi,\: P\in S_{\text{ECC}}^\bot\setminus S_{\text{ECC}},\: Q\in\{I,Z\}^{\otimes n},\: R\in\{I,X\}^{\otimes n}\} \right|}{\left|\Pi_{3n}\right|}. \end{equation}
We can split the set of permutations as the intersection of just the error correction and the well-studied trap detection probability
\begin{equation}\begin{split}
\Pi^0(E) &:= \{\pi\in\Pi_{3n}\colon E=\pi^\dag(P\otimes T)\pi,\: P\in S_{\text{ECC}}^\bot\setminus S_{\text{ECC}},\: T\in\mathcal{G}_{2n}\},\\
\Pi^1(E) &:= \{\pi\in\Pi_{3n}\colon E=\pi^\dag(P\otimes Q\otimes R)\pi,\: P\in \mathcal{G}_n,\: Q\in\{I,Z\}^{\otimes n},\: R\in\{I,X\}^{\otimes n}\},
\end{split}\end{equation}
such that the purity testing is upper bounded by the minimum size of both sets
\begin{equation}\label{eq:fix_purity} \Pr_{k\in\mathcal{K}}\left(E\in S_k^\bot\setminus S_k\right)\leq\min\left\{\frac{\left|\Pi^0(E)\right|}{\left|\Pi_{3n}\right|},\frac{\left|\Pi^1(E)\right|}{\left|\Pi_{3n}\right|}\right\}. \end{equation}
The second term is bounded by $(2/3)^{\omega/2}$, see~\cite{broadbent_quantum_2013}. For the first term, recall that concatenated error-correcting codes correct \emph{typical} errors, see~\Cref{sec:error}. Although here the bound is for i.i.d.\ Pauli noise, we can use this to bound the probability of our fixed weight attack. Note that the set of Pauli operators acting non-trivially and being undetected is exactly the set of Pauli operators that the error-correcting code fails to decode correctly, and we can write this in terms of random variables. Let us define the following set of i.i.d.\ random variables $X_1,\ldots, X_{3n}$ such that $X_j\in\{X,Y,Z\}$ with probability $\omega/3n$ and $X_j=I$ with probability $1-\omega/3n$. Let us denote by $\textbf{X}$ the tensor product of the first $n$ variables, $\textbf{X}:=X_1\otimes\cdots\otimes X_n$ as in~\Cref{sec:error}. Then the weight of these variables is
\begin{equation}
    \omega(X_j):=
    \begin{cases} 1 & \text{if $X_j\not=I$}\\
    0 & \text{otherwise}  \end{cases},
    \quad\text{with}\quad \Pr(\omega(X_j)=1)=\frac{\omega}{3n}.
\end{equation}
Now if we condition on a fixed amount of registers suffering an error, we have the bound
\[ \frac{\left|\Pi^0(E)\right|}{\left|\Pi_{3n}\right|}\leq \Pr(\textbf{X}\in S_{\text{ECC}}^\bot\setminus S_{\text{ECC}}\left|\sum_{j=1}^{3n} \omega(X_j)=\omega\right.)\leq\frac{\Pr\left(\textbf{X}\in S_{\text{ECC}}^\bot\setminus S_{\text{ECC}}\right)}{\Pr(\sum_{j=1}^{3n} \omega(X_j)=\omega)}. \]
The numerator in the LHS is exactly the probability of failed recovery of an error-correcting code when the message is sent through a channel with error probability $\omega/{3n}$ per qubit, which is bounded by~\cref{eq:threshold}. The denominator is just the probability of the binomial distribution having the expected value and is lower bounded by 
\[ \Pr(\sum_{j=1}^{3n} \omega(X_j)=\omega)\geq\frac{9}{10}\frac{1}{\sqrt{2\pi\omega(1-\omega/3n)}}, \]
see~\Cref{app:binom}. Therefore, 
\begin{equation}\label{eq:ecc_bound} \frac{\left|\Pi^0(E)\right|}{\left|\Pi_{3n}\right|}\leq \pth \frac{9(\omega/3n\pth)^{(t_1+1)^M}}{10\sqrt{2\pi\omega(1-\omega/3n)}}, \end{equation}
where $\pth=\binom{n_1}{t_1+1}^{-1}$ is the threshold of the error-correcting code. We can make the above bound as small as desired by increasing the levels $M$ of concatenation whenever $\omega<3n\pth$.

Of course the attacker will choose the best possible attack for a given protocol, thus we need a bound independent of the weight $\omega$. Since the error-correcting code will always correct attacks of weight smaller than $t:=\frac{d_1^M-1}{2}$, the bound in~\cref{eq:ecc_bound} is only relevant for $\omega\geq t$. By taking the maximum over the possible weights in~\cref{eq:fix_purity} we obtain the following purity-testing parameter. For an arbitrary Pauli error $E\in\mathcal{G}_{3n}$, the trap code is purity testing with
\begin{equation}\label{eq:min_ecc_dec} \Pr_{k\in\mathcal{K}}\left(E\in S_k^\bot\setminus S_k\right)\leq\max_{\substack{\omega\geq t}}\left\{\min\left\{(2/3)^{\omega/2},\pth \frac{9(\omega/3n\pth)^{(t_1+1)^M}}{10\sqrt{2\pi\omega(1-\omega/3n)}}\right\}\right\}. \end{equation}
\end{proof}
\end{prop}

While the trap detection is a decreasing function on $\omega$, independent of the total size of the protocol, the probability of the error correction failing is an increasing function on $\omega$, and decreasing in $M$ only when $\omega< 3n\pth$, which depends on the size $n$. In orther to study this intersection we can divide different cases:
\begin{itemize}
    \item Case 1: $t>3n\pth$. Which will not be the case for $M$ big enough, but here the error correction does not play a role, thus the security is the maximum of the detection probability which will be $(2/3)^{t/2}$. By a first order approximation, the amount of errors that an error-correcting code corrects grows as  
    \[ t\simeq n^{\log_{(n_1)}(d_1)}, \]
    where the size $n_1$ and distance $d_1$ depend on the error-correcting code. In particular, for the Steane code $\log_7(3)\simeq 0.56$ and for the $[[5,1,3]]$-code $\log_5(3)\simeq 0.68$.
    \item Case 2: $t<3n\pth$. Which is the case for $M$ large enough, and here the error correction plays a role. In order to improve the exponential decay of the error correcting code, we need the intersection to occur in a weight $\omega = n^{\alpha}$ for some $\alpha>\log_{(n_1)}(d_1)$. However, if we substitute this value of $\omega$ for both functions in~\cref{eq:min_ecc_dec} and take logarithms in the same bases we obtain 
    \[ n^\alpha/2=\log_{(2/3)}(\frac{9\pth}{10\sqrt{2\pi}})+\log_{(2/3)}(n^{-\alpha/2}(1-n^{\alpha-1}/3)^{-1/2})+(t_1+1)^M\log_{(2/3)}(n^{(\alpha-1)}/3\pth), \]
    which allows us to compute $\alpha$ for large enough concatenations $M$,
    \[\alpha\simeq \log_{n_1}(t_1+1),\]
    which for the Steane code is $\log_7(2)\simeq 0.35$ and for the $[[5,1,3]]$-code $\log_5(2)\simeq0.43$.
\end{itemize}
Observe how the scaling of the weight of intersecting point is worse than the scaling of the errors, which means that for enough concatenations $M$ of the error-correcting code, the security of the trap code will be determined by the probability of the traps detecting an error. We can conclude that Broadbent et al.'s~\cite{broadbent_quantum_2013} security analysis is a fair comparison. 

\end{document}